\documentstyle[12pt,epsf]{article}
\newcommand{\be}{\begin{equation}}
\newcommand{\bea}{\begin{eqnarray}}
\newcommand{\eea}{\end{eqnarray}}
\newcommand{\beas}{\begin{eqnarray*}}
\newcommand{\eeas}{\end{eqnarray*}}
\newcommand{\ba}{\begin{array}}
\newcommand{\ea}{\end{array}}
\newcommand{\ee}{\end{equation}}

\newcommand{\tr}{{\rm Tr}\ }
\newcommand{\nbox}{{\,\lower0.9pt\vbox{\hrule \hbox{\vrule height 0.2 cm \hskip
0.2 cm \vrule height 0.2 cm}\hrule}\,}}

\makeatletter
\@addtoreset{equation}{section}
\makeatother

\renewcommand{\a}{\alpha}

\newcommand{\pa}{\partial}

\newcommand\text[1]{\rm #1}

\newcommand{\CC}{\hbox{\xiiss C\kern-.4emI}}
\newcommand{\RR}{\hbox{\xiiss R\kern-.45emI}}
\newcommand{\ZZ}{\hbox{\xiiss Z\kern-.4emZ}}
\newcommand{\CCs}{\hbox{\ixss C\kern-.4emI}}
\newcommand{\ZZs}{\hbox{\ixss Z\kern-.4emZ}}
\newcommand{\pasl}{\pa\kern-.55em /}
\def\href#1#2{#2}

\def\a{\alpha}
\begin{document}
\begin{titlepage}
\hfill
\vbox{
    \halign{#\hfil         \cr
           hep-th/0205185 \cr
           SU-ITP-02/14 \cr
           } % end of \halign
      }  % end of \vbox
\vspace*{20mm}
\begin{center}
{\Large {\bf Matrix Perturbation Theory \\  For M-Theory On a PP-Wave}\\ }

\vspace*{15mm}
\vspace*{1mm}
%{Authors}
{K. Dasgupta, M. M. Sheikh-Jabbari, M. Van Raamsdonk} 

\vspace*{1cm}

{Department of Physics, Stanford University\\
382 via Pueblo Mall, Stanford CA 94305-4060, USA\\
{\it keshav, jabbari, mav @itp.stanford.edu} }

\vspace*{1cm}
%%\maketitle
\end{center}

\begin{abstract}

In this paper, we study the matrix model proposed by Berenstein, Maldacena, and 
Nastase to describe M-theory on the maximally supersymmetric pp-wave. We show 
that the model may be derived directly as a discretized theory of 
supermembranes 
in the pp-wave background, or alternatively, from the dynamics of D0-branes in 
type IIA string theory. We consider expanding the model about each of its 
classical supersymmetric vacua and note that for large values of the mass 
parameter $\mu$, interaction terms are suppressed by powers of $\mu^{-1}$, so 
that the model may be studied in perturbation theory. We compute the exact 
spectrum about each of the vacua in the large $\mu$ limit and find the 
complete (infinite) set of BPS states, which includes states preserving 2, 4, 
6, 8, or 16 supercharges. Through explicit perturbative calculations, we then 
determine the effective coupling that controls the perturbation expansion for 
large $\mu$ and estimate the range of parameters and energies for which 
perturbation theory is valid. 

\end{abstract}

\end{titlepage}

\tableofcontents
\vskip 1cm
%\newpage
\section{Introduction}

At the present time, the most promising candidate for a microscopic definition 
of eleven dimensional M-theory is the large $N$ limit of the quantum mechanics 
obtained from dimensionally reducing ten dimensional $U(N)$ SYM theory to 0+1
dimensions, with Hamiltonian
\be
\label{h1}
H_0 = R\ \tr \left( {1 \over 2} \Pi_A^2 - {1 \over 4} [X_A, X_B]^2 
- {1 \over 2} \Psi^\top \gamma^A [X_A, \Psi] \right) \; \ \ A,B=1,\cdots , 9.
\ee

The initial evidence for this suggestion came from the idea that the quantum 
completion of eleven dimensional supergravity should be realized as a theory 
of quantum supermembranes just as supergravity in ten dimensions is given its 
quantum completion via superstrings. Unlike the superstring case, the 
worldvolume supermembrane theory is an interacting theory and the theory 
contains no tunable coupling constant; however, de Wit, Hoppe, and Nicolai 
\cite{dhn} showed that in the light cone gauge, the supermembrane action can be 
regularized to give the quantum mechanics (\ref{h1}), providing a possible 
avenue for directly studying the quantum supermembrane. It was soon realized 
that in contrast to the discrete spectrum of states obtained for a quantum 
superstring, the spectrum of the matrix model (\ref{h1}) is
continuous, and at the time this appeared to be a rather troubling feature of 
the quantum supermembrane. 

The matrix model was given new life following the realization that a quantum 
completion of eleven-dimensional supergravity  should arise directly from 
string theory as the strong coupling limit of type IIA strings. 
Based on this connection, Banks, Fischler, Shenker, and Susskind (BFSS) 
\cite{bfss} argued that M-theory in the ``infinite momentum frame'' should be 
described by the limit 
of a large number of D0-branes in type IIA string theory at low energies and 
weak coupling. Remarkably, this system is described again by the quantum 
mechanics (\ref{h1}), so essentially the same conclusion is reached from a 
very different approach. Happily, the D0-brane point of view provided a 
very natural explanation for the continuous spectrum of the matrix model: the 
theory should be interpreted as a second quantized theory containing both 
single and multiparticle states, rather than a first quantized theory. This 
point of view also provided a precise interpretation for the finite $N$ 
matrix model, as the DLCQ of M-theory with $N$ units of momentum 
in the compact direction \cite{susskind}.

Since the work of BFSS, matrix theory has passed numerous tests (for a recent 
review, see \cite{taylor}), however an 
honest quantum mechanical study of the model seems extremely difficult, due to
the flat directions in the potential (resulting in the continuous spectrum 
mentioned earlier), the difficulty of distinguishing single-particle and 
multi-particle states, the lack of a tunable coupling constant (interactions 
are of order one), and the large number of degrees of freedom in the large $N$
 limit. Indeed, even determining the normalizable ground state wavefunction 
for the case $N=2$ (proven to exist in \cite{ss}) is an unsolved and 
notoriously difficult problem. 

Recently \cite{bmn}, it has been shown by Berenstein, Maldacena, and Nastase 
(BMN) that the matrix model (\ref{h1}) is actually part 
of a one-parameter family of matrix models with 32 supercharges given by
\bea\label{PPmatrix}
H = H_0 &+& {R \over 2} \tr \Big(\sum_{i=1}^3 {\left({\mu\over 
3R}\right)^2 } X_i^2 + \sum_{a=4}^9 \left({\mu \over 6R}\right)^2 X_a^2 
\cr
&& \qquad \qquad + 
i {\mu \over 4R} \Psi^\top \gamma^{123} \Psi  + i {2\mu \over 3R} 
\epsilon^{ijk} X_i X_j X_k \Big)\ .   
\eea
The matrix model for general $\mu$ was proposed by BMN to give a 
description of the DLCQ of M-theory on the maximally supersymmetric 
pp-wave background \cite{penrose,guven,kg,blau}  of eleven-dimensional 
supergravity\footnote{We have used conventions in which $x^{\pm} = (t \pm 
x)/\sqrt{2}$, different from those in BMN.} 
\bea\label{PPwave}
ds^2 &=& - 2 dx^+ dx^- + \sum_{A=1}^9 dx^A dx^A - \left(\sum_{i=1}^3 {\mu^2 
\over 9} x^i x^i + \sum_{a=4}^9 {\mu^2 \over 36} x^a x^a\right) dx^+ dx^+\cr
F_{123+} &=& \mu
\eea
In the large $N$ limit with fixed $p^+ = {N \over R}$ and fixed $\mu$, this 
matrix model should then describe M-theory on the pp-wave background in a 
sector with boost-invariant momentum parameter $\mu p^+$. 

As pointed out in \cite{bmn}, the pp-wave matrix model for non-zero 
$\mu$ avoids a number of difficulties associated with the $\mu=0$ case. 
Firstly, the quadratic terms in the potential remove all flat directions, and 
yield a potential that becomes infinite in all directions. This implies a 
discrete spectrum for the massive matrix model and, as shown in \cite{bmn}, 
an isolated set of classical supersymmetric vacua. These vacua may be 
associated with all possible ways of dividing up the momentum into different 
numbers of membranes, from a single membrane carrying $N$ units of momenta to 
$N$ membranes each carrying a single unit of momentum. Since these vacua are 
separated by potential barriers, we see that the massive matrix model (at low 
energies) also has a natural distinction between single and multi-membrane 
states. 

In this paper, we study the massive matrix model and explore the consequences 
of these and a further simplifying feature: that
the existence of a tunable parameter in the matrix model opens up the 
possibility of a weakly coupled regime which can be studied in perturbation 
theory. 

We now provide a concise summary of our results as we outline the remainder of 
the paper.

Before proceeding to study the model we first clarify its origins.\footnote{The 
derivation provided in \cite{bmn} was somewhat 
indirect, showing that the $N=1$ model matches the action for a superparticle 
in the pp-wave background and then writing an extension of the theory to 
general $N$ consistent with supersymmetry.} In section 2 we show
that the massive matrix model arises directly as the regularized theory of 
light-cone gauge supermembranes in the maximally supersymmetric pp-wave 
background. In section 3, we note that the model may also be derived from the 
D0-brane perspective, although here one has to assume that the weak background 
D0-brane action continues to be valid even in a region of high curvature, 
perhaps a consequence of supersymmetry. 

In section 4, we proceed with a direct study of the massive matrix 
model. We begin by reviewing the 
action and its symmetries and identify explicit matrix theory expressions for 
the generators of the superalgebra, verifying that the (anti)commutation 
relations are satisfied correctly. We note that the appearance of rotation 
generators along with the Hamiltonian in the anticommutator of 
supersymmetry 
generators allows BPS states carrying angular momentum.\footnote{An example of 
such a state has been given in \cite{bak}.} After these 
generalities, we review the classical vacua of the theory, collections of 
fuzzy spheres corresponding to sets of membrane giant gravitons \cite{myers, 
mst}  
carrying various fractions of the total light-cone momentum, $p^+$. We 
expand the matrix model 
about each of these classical vacua and note that in each case we obtain a 
quadratic theory with interactions suppressed by a coupling $(R/ \mu)^{3 
\over 2}$. We conclude that for fixed values of the other parameters, the 
model may be studied in perturbation theory for sufficiently large $\mu$. 

In section 5, we consider the theory in the $\mu \to \infty$. Here, the 
different vacua become superselection sectors separated by an infinite 
potential barrier and interactions vanish. For each vacuum, we explicitly 
diagonalize the quadratic action, obtaining the precise oscillator masses and 
thus the exact spectrum of the theory. We find that the oscillator masses are 
proportional to $\mu$ and completely independent of $N$. In fact, $N$ 
plays the role of an ultraviolet cutoff leaving a finite tower of oscillator 
masses that becomes an infinite tower for $N \to \infty$. This fits very well 
with the regularized supermembrane picture, and in particular, we may interpret 
the large $N$ limit of the spectrum obtained from the single fuzzy sphere vacuum 
as the quadratic spectrum of a single supermembrane in a large $\mu
p^+$ limit of the pp wave background. From the exact spectrum, we are
able to explicitly identify all 
BPS states of the theory and find infinite towers of states preserving 2, 
4, 6, and 8 supercharges in addition to the 1/2 BPS vacuum states.

In section 6, we move away from the $\mu = \infty$ limit and investigate 
perturbation theory about the quadratic actions. We expand the interaction 
Hamiltonian in terms of the canonically normalized oscillators, and find that 
the parameter suppressing the interaction terms is different for different 
vacua. In particular, the coupling for the $N$-membrane vacuum remains $g = 
(R/ \mu)^{3 \over 2}$ while the coupling for the single-membrane vacuum is 
reduced to $g = (R/( \mu N))^{3 \over 2}$. This latter quantity, equals to 
$r^{-3 / 2}$ where $r$ is the radius of the membrane sphere, remains finite in the M-
theory limit of fixed $\mu$ and fixed $p^+ = N/R$, giving the coupling constant 
for the continuum field theory living on the single spherical 
supermembrane. 

Since it is often the case that bare couplings become modified to some effective 
coupling in theories with many degrees of freedom, we then compute explicitly 
the energy shifts for certain non-BPS excited states of the single-membrane and 
$N$-membrane vacua. In both cases, we find that the leading 
corrections to the energy are suppressed by an effective coupling 
$\lambda = g^2 K$ relative to the energy in the free theory, where $g$ is 
the bare 
coupling for the corresponding vacuum and $K$ is the multiplicity of the 
representation (number of coincident branes). Thus, the leading perturbative 
correction in the irreducible vacuum is small when
$g^2 = 1 / (\mu p^+)^3 \ll 1$ while a small leading correction for the $X=0$ 
vacuum requires a much stronger condition, $N^4 / (\mu p^+)^3 \ll 1$. 

For fixed values of the parameters, we then estimate the energy scale at which 
perturbation theory will break down (i.e. the interaction Hamiltonian becomes of 
the same order of magnitude as the quadratic piece). For generic directions away 
from the single-membrane vacuum, we find that interactions become important at 
an energy $E/\mu \sim 1/g^2 = (\mu N / R)^3$ but there are certain directions 
(corresponding to the large membrane shrinking and a small membrane growing at 
the origin) for which interactions become important at $E/ \mu \sim 1 / (g^2 N) 
= {1 \over N} (\mu N / R)^3 $. We expect that a valid perturbative regime will 
certainly exist when this energy is significantly greater than the energies of 
the lowest excited states, $g^2 N \ll 1$, as depicted in figure 1. On the 
other 
hand, our direct perturbative calculations indicate small perturbative 
corrections for low energy states under the much weaker condition $g^2 \ll 1$.

In section 7, we discuss the implications of our results for the large $N$ 
limit of the model. Our calculation of the leading energy correction indicates 
that perturbation theory near the single membrane vacuum may be valid even in 
the large $N$ limit as long as $\mu p^+ \gg 1$. On the other hand, the more 
conservative estimate based on the height of the barrier between vacua suggests 
that we always exit the perturbative regime in the large $N$ limit defining M-
theory (as in figure 1). In any case, our results explicitly demonstrate that in 
this limit, the theory develops finite zero-energy flat directions corresponding 
to decreasing the radius of any membrane while increasing the radius of any 
other membrane (or nucleating a new spherical membrane  at the origin). This 
indicates the existence of massless singular BPS domain walls, or tensionless 
strings, in the supermembrane world-volume field theory. 

Finally, we offer some concluding remarks in section 8 and technical 
results in a set of appendices.

\begin{figure}
\centerline{\epsfysize=2.0truein \epsfbox{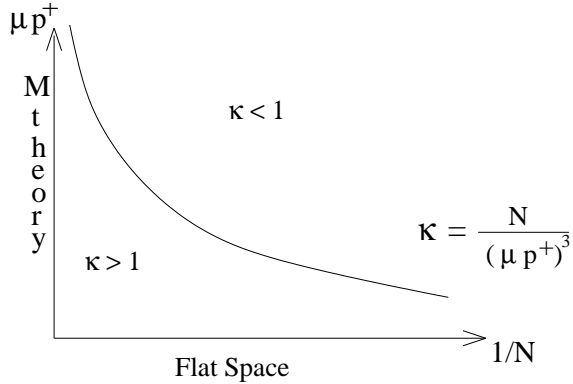}}
%\vskip -0.4 cm
 \caption{Regime of validity for perturbation theory (conservative)}
\label{pert2}
\end{figure}

\section{The matrix model from supermembranes}

It is well known that the usual Matrix theory action can derived
as a discretized version of the light-cone gauge supermembrane action
in flat space, 
where Poisson brackets on the membrane worldvolume are replaced with
matrix commutators \cite{dhn}. In other words, the 
invariance of the supermembrane action under area preserving 
diffeomorphisms is translated to the $U(N), N\to\infty$  gauge 
invariance of the dimensionally reduced SYM theory which defines the matrix 
model.

In this section we show that the same story applies in the case of the
pp-wave background. That is, a discretized version of the light-cone 
supermembrane action on the maximally supersymmetric eleven dimensional 
pp-wave background exactly reproduces the matrix model proposed by BMN. 

To start with, we borrow  the supermembrane action from \cite{dpp}:
\be\label{membact}
S[Z(\sigma)] = \int d^3 \sigma \; \sqrt{- G(Z(\sigma))}+
\int C  \ ,
\ee
where $Z^M(\sigma) = X^{\mu}(\sigma), \theta^{\alpha}(\sigma)$ are the 
superspace embedding coordinates and $\sigma^i, i= 0,1,2$ are the 
world-volume membrane coordinates. Here,
\be\label{GGG}
G(Z(\sigma))={\rm det} (\Pi_i^{\mu}\Pi_j^{\nu}G_{\mu\nu})\ , 
\ee
where $\Pi^M_j = {\partial Z^N \over 
\partial \sigma^j} E^M_N$ is the pull back of the supervielbein
to the membrane world-volume and $G_{\mu\nu}$ is the background
metric, which in our case is the 
pp-wave metric (\ref{PPwave}). The three form $C$ field is
\be\label{CCC}
C= {1\over 6} e^r \wedge e^s \wedge e^t ~C_{rst}+ C_F\ ,
\ee
where $e^r=e^r_{\mu}dX^{\mu}$ and $e^r_{\mu}$ are the vielbeins. The $C_F$ 
term, which denotes the fermionic contributions to the $C$ superfield, will be
presented explicitly in section 2.2.

We begin with the  bosonic terms and then consider the fermionic terms in 
the supermembrane action.

\subsection{Bosonic terms}

In order to obtain the matrix model from the membrane action one should fix 
the light-cone gauge.  
For that it will be convenient to use the coordinates 
$(X^+, X^-, X^A)$ with $A = 1,...,9$. The light-cone condition 
is then $X^+ = \tau$, while the other worldvolume coordinates are 
denoted by $\sigma^r,\ r=1,2$.  One should note that to fix the light-cone 
gauge consistently we should also impose  the following constraint on the 
phase-space variables
\be
P_A \partial_r X^A + P_- \partial_r X^- = 0 \ ,
\ee
where $P_A, P_-$ are the momentum conjugate to $X^A$ and $X^-$, respectively.

Upon fixing the light-cone gauge, the bosonic part of the total
Hamiltonian becomes \cite{dpp}
\bea\label{hamil}
H_{bos}&=&\int d^2\sigma {\Big (} {G_{+-} \over P_- - C_-} {\Big [}
{1\over 2} (P_A - C_A - {P_- - C_- \over G_{+-}} G_{A+} )^2 + {1\over 4}
(\epsilon^{rs} \partial_r X^A \partial_s X^B)^2 {\Big ]} \cr
&-& \! \! \! \! \! {P_- - C_- \over 2 G_{+-}}G_{++} - C_+ + {1\over P_- - C_-}
{\Big [} \epsilon^{rs} \partial_r X^A \partial_s X^B P_A C_{+-B} + 
C_- C_{+-} {\Big ]} {\Big )} 
\eea
where 
\bea
C_A &=& -\epsilon^{rs} \partial_r X^- \partial_s X^B C_{-AB} 
+ {1\over 2}\epsilon^{rs} \partial_r X^B \partial_s X^C C_{ABC}\  \cr
C_{\pm} &=& {1\over 2} \epsilon^{rs} \partial_r X^A \partial_s X^B 
C_{\pm AB}\  \\ 
C_{+-} &=& \epsilon^{rs} \partial_r X^- \partial_s X^A C_{+-A}\ .\nonumber 
\eea
Now let us concentrate on the pp-wave background, where
\bea\label{bg}
G_{--} &=& G_{A-} = 0, ~~~ G_{+-} = -1, ~~~ G_{AB} = \delta_{AB}\cr
G_{++}& = &- F^2\equiv -({\mu^2\over 9} \sum_{i=1}^3 X_i^2+{\mu^2\over 
36} \sum_{a=1}^6 X_a^2\ ) ,
\eea
and 
\be\label{CC}
C_-=C_{+-}=C_A=0,\  C_+={1\over 2}\epsilon^{rs}
\partial_r X^A \partial_s X^B C_{+AB}=
{\mu\over 3!}\epsilon_{ijk}\epsilon^{rs}\partial_r X^i \partial_s X^j X^k\ .
\ee
In this background the light-cone Hamiltonian  above  simplifies significantly 
and we are left with
\bea\label{PPmembrane}
H_{bos} &=& {1\over p^+}\int d^2\sigma {\Bigg[}  {1 \over 2}P_A^2+
{1\over 4}\{X^i, X^j\}^2+ {1\over 4}\{X^a, X^b\}^2 +{1\over 2}\{X^i, 
X^a\}^2\cr 
&&\ \ \ \ \ \
+{1\over 2}\left(({\mu p^+\over 3})^2 {(X^i)}^2+
({\mu p^+\over 6})^2 {(X^a)}^2\right)-
{\mu p^+\over 6} \epsilon^{ijk} \{X^i, X^j\} X^k{\Bigg ]} .
\eea
where $i,j=1,2,3, \ a,b=1,2,\cdots , 6$,   $p^+=-P_-$ and  
$$
\epsilon^{rs} \partial_r X^A \partial_s X^B=\{ X^A,X^B\}\ . 
$$

\subsection{Fermionic terms}

To get the fermionic terms we have to study the supervielbein and the tensor 
field backgrounds to all orders in $\theta$. For coset spaces (of
which the pp-wave is an example), the exact expressions for the 
supervielbein $E_N^B$,\footnote{The $B = (r,a)$ indices are tangent
space indices and the curved 
indices are $N = (\mu, \alpha)$, where the first entries are bosonic and 
the 
second entries are fermionic.} in terms of the component worldvolume
and background fields have been worked out in \cite{dpps}, and are
given by\footnote{The coset space approach was also used in \cite{metsaev, 
mt} to derive the action for light cone type IIB string theory in the pp-wave 
background.}
\bea\label{ebg}
E &=& D\theta +\sum_{n=1}^{16}~ {1\over (2n+1)!} {\cal M}^{2n} 
D\theta \\
E^r &=& e^r + \bar{\theta} \Gamma^r D\theta +2 \sum_{n=1}^{15}~ 
{1\over (2n+2)!} {\bar \theta} \Gamma^r {\cal M}^{2n} D\theta\ .
\eea
In these expressions,
\bea
e^r &=& e^r_\mu dX^\mu  \\
\omega^{rs}&=&\omega^{rs}_\mu dX^\mu \\ 
D\theta &\equiv& d\theta + e^r T^{stuv}_r \theta F_{stuv} - {1\over 4} 
\omega^{rs} \Gamma_{rs} \theta\ ,
\eea
where $\omega^{rs}$ is the spin connections, $F_{stuv}$ is the field 
strength of the three form $C$ and 
\be\label{tdef}
 T_{r}^{\ stuv}= {1\over 2!3!4!} {\Big (} 
\Gamma_{r}^{\ stuv}- 8 \delta^{\ [s}_{r}\Gamma^{tuv]} {\Big )}\ .
\ee
Finally, the matrix ${\cal M}^2$ is defined as
\be\label{mmat}
({\cal M}^2)^a_b = 2 (T^{stuv}_r \theta)^a F_{stuv} ({\bar \theta} \Gamma^r)_b 
-{1\over 288} (\Gamma_{rs} \theta)^a {\Big (}{\bar \theta} [
\Gamma^{rstuvw} F_{tuvw} +24 \Gamma_{tu}F^{rstu}] {\Big )}_b\ .
\ee
%\bea\label{viel}
%&& E_{\mu}^r = e^r_{\mu} + 
%2 {\bar \theta} \Gamma^r \psi_{\mu} +
%{\bar \theta} \Gamma^r {\Big [} -{1\over 4} \omega^{st}_{\mu} \Gamma_{st}
%+ T^{\nu\rho\sigma\lambda}_{\mu} F_{\nu\rho\sigma\lambda}{\Big ]} \theta 
%\cr && E_{\mu}^a = \psi_{\mu}^a - {1\over 
%4}\omega^{rs}_{\mu}(\Gamma_{rs}\theta)^a
%+ (T^{\nu\rho\sigma\lambda}_{\mu}\theta)^a F_{\nu\rho\sigma\lambda} \cr
%&& E_{\alpha}^r = -({\bar \theta} \Gamma^r)_{\alpha} \cr
%&& E_{\alpha}^a = \delta^a_{\alpha}  
%\ee
In the pp-wave background,
\be
e^-_{\ -} = e^+_{\ +} = 1\ ,\ \ e^A_{\ B} = \delta^A_{\ B}, ~~~e^-_{\ +} = 
{F^2\over 2}\ , 
\ee
and all the other components are zero. Then, the only non-vanishing 
component of the spin connection is
\be\label{spincon}
\omega^{-A}_+ = {1\over 2} \partial_A F^2\ .
\ee

To work out the explict form of the fermionic fields we should choose a 
specific representation for $\Gamma$ matrices. The $32 \times 32$ 
$\Gamma$ matrices can be written as 
\be
{\Gamma^+}=\sqrt{2}
\left(\matrix{
 0 & 0  \cr
     i{\bf 1_{16}}  & 0
}\right)\ \ ,~~~
\Gamma^- =\sqrt{2}
\left(\matrix{
 0 &  i{\bf 1_{16}}  \cr
     0  &    0
}\right)\ \ ,~~~
{\Gamma^A}=
\left(\matrix{
 \gamma^A & 0  \cr
     0  & -\gamma^A
}\right)\ \ 
\ee
where the $\gamma^A$ are $16 \times 16$ Euclidean $SO(9)$ gamma matrices. 
\footnote{Note that $\Gamma^{\pm}={1\over \sqrt{2}}(\Gamma^0\pm \Gamma^{10})$, 
and $\bar\theta=i\theta^{\dagger}\Gamma^0$.} 
%Also note that ${(\Gamma^+)}^2={(\Gamma^-)}^2=0$.}
 We can also use $\kappa$-symmetry to fix the (light-cone) gauge:
\be\label{gauge}
\Gamma^+ \theta = 0\ .
\ee
In this gauge we have the advantage of significant simplifications
such as
\be\label{Gauge}
\bar\theta\Gamma^+=\bar\theta\Gamma^{+-}\partial\theta= 
\bar\theta\Gamma^{+A}\partial\theta=\bar\theta\Gamma^{+AB}\partial\theta=
\bar\theta\Gamma^{A}\partial\theta=\bar\theta\Gamma^{AB}\partial\theta=0, 
\cdots 
\ee
One can solve the gauge condition (\ref{gauge}) explicitly:
\be\label{gaugepsi}
\theta={1\over 2^{1/4}} \left(\matrix{0  \cr \Psi }\right)\ , \ \ \  
\bar\theta={1\over 2^{1/4}}\left(-\Psi^{\top}\ \ \ \ 0 \right)\ , 
\ee
where $\Psi$ is a 16 dimensional (Majorana) spinor. The 
normalization for $\Psi$ have been chosen so that
\[
i\bar\theta \Gamma^-\theta=\Psi^\top \Psi\ .
\]

To work out the action explicitly, we first note that for the pp-wave
case, as discussed earlier, 
the only non-trivial background field component appearing in ${\cal M}^2$ is 
$F_{+123} =-F^{-123}$. Therefore, using the gauge condition (\ref{gauge}) one 
can 
show that 
\be
{\cal M}^2 = 0\ .
\ee
That is, the only non-vanishing fermionic contributions come from the 
expansion of the supervielbeins to the leading order in $\theta$.

Now let us focus on the fermionic contributions coming from the 
$\sqrt{-G}$ terms in the action. For these, we need to work out 
$\Pi_j^r$'s:
\be
\Pi_i^r=E^r_{\nu}\partial_i X^{\nu}+E^r_{\alpha}\partial_i \theta^\alpha\ .
\ee 
Using the Eq.(\ref{ebg}), we have
\bea
E^r_{\nu}&=&e^r_{\nu}+\bar\theta\Gamma^r D_{\nu}\theta, \cr
E^r_{\alpha}&=&\bar\theta\Gamma^r D_{\alpha}\theta\ ,
\eea
where
\[
\bar\theta\Gamma^r D_{\nu}\theta=
{i\mu\over 4}\delta^+_{\nu}\delta^r_-\ \Psi^{\top}\gamma^{123}\Psi\ ,
\] 
and $E^r_{\alpha}=-(\bar\theta \Gamma^r)_{\alpha}$. Putting this 
all together we obtain
\bea\label{Pi}
\Pi_i^-&=&\partial_i X^- +\partial_i X^+({F^2\over 2}+
{i\mu\over 4}\Psi^{\top}\gamma^{123}\Psi)+i \Psi^{\top}\partial_i\Psi\ , 
\cr 
\Pi_i^+&=&\partial_i X^+, \ \ \ \Pi_i^A=\partial_i X^A\ .
\eea
We can see that all the effects of the pp-wave background appear in
 the $\partial_i X^+$ factor of $\Pi_i^-$. Hence, in the membrane 
Hamiltonian (\ref{hamil}) $G_{++}$ should be shifted to 
\[
G_{++}+{i\mu\over 2}\Psi^{\top}\gamma^{123}\Psi \; ,
\]
which leads to a ``mass term'' for the fermions. Note that the other fermionic 
term $i \Psi^{\top}\partial_i\Psi$ will not contribute to the 
light-cone Hamiltonian (because it involves first order time derivatives).

The only remaining terms to be considered are the fermionic contributions to 
the $C$ term, $C_F$ \cite{dpps}:
\be
C_F=- \int _0^1 tdt~ {\bar \theta} \Gamma_{rs} ~E(t) \wedge E^r(t) \wedge 
E^s(t)\ , 
\ee 
where $t$ is an auxiliary parameter and $E(t)$, $E^r(t)$ are obtained
from (\ref{ebg}) by the shift $\theta\to t\theta$. As we 
discussed, using the gauge  fixing condition (\ref{gauge}), ${\cal M}^2$ 
is zero and hence
\[
C_F=- \int _0^1 t^2dt~ {\bar \theta} \Gamma_{rs}D\theta \wedge 
(e^r+ t^2 \bar\theta \Gamma^r D\theta)\wedge 
(e^s+ t^2 \bar\theta \Gamma^s D\theta)\ . 
\]
In the gauge (\ref{gauge}), ${\bar \theta} \Gamma_{rs}D\theta$ is 
zero unless $\{rs\} = \{+A\}$, while $\bar\theta \Gamma^r D\theta$ is
zero if $r$ is $+$. Therefore
the only surviving term is 
\bea
C_F&=& - \int _0^1 t^2dt~ {\bar \theta} \Gamma_{rs}D\theta\wedge e^r\wedge 
e^s
\cr
&=& {1\over 3} \epsilon^{ijk} \bar\theta\Gamma^{-A}
\partial_i X^+ \partial_j \theta \partial_k X^A \cr
&=&{i\over 2}\Psi^{\top}\gamma^A\{X^A,\Psi\}\ .
\eea

In summary, gathering together all fermionic and bosonic terms, we have shown 
that the total light-cone gauge Hamiltonian for membranes in the maximally
supersymmetric pp-wave background, with properly fixed $\kappa$-symmetry, is:
\bea\label{membranetotal}
H_{total} = \int \!\!\! && d^2\sigma {\Bigg\{} {1\over p^+} {\Bigg [} {1 
\over 
2}P_A^2+{1\over 4}
\{X^A, X^B\}^2 +{1\over 2}\left(({\mu p^+\over 3})^2 {(X^i)}^2+
({\mu p^+\over 6})^2 {(X^a)}^2\right)\cr &&
\ \ \ \ -{\mu p^+\over 6} \epsilon^{ijk} \{X^i, X^j\} X^k{\Bigg ]}
-{i\over 2}\Psi^{\top}\gamma^A\{X^A,\Psi\}-{i\over 
4}\mu\Psi^{\top}\gamma^{123}\Psi{\Bigg\}}\ . 
\eea

The matrix model Hamiltonian proposed by BMN now follows immediately by
applying the regularization prescription of \cite{hoppe,dhn}, in which we 
replace
functions on the membrane worldvolume by hermitian $N \times N$ matrices,
integrals with traces, and Poisson-brackets with commutators, i.e.
\bea
X^A(\sigma)&\longleftrightarrow & {1\over N} X^A\cr 
\Psi(\sigma)&\longleftrightarrow & \Psi \cr 
p^+\int d^2\sigma &\longleftrightarrow & {1\over R} Tr \cr 
\{\ , \ \}_{PB} &\longleftrightarrow & -i\ [\ ,\ ]\ . \nonumber 
\eea

In the next section, we show explicitly how the matrix model may also be 
obtained from the D0-brane perspective.

\section{The matrix model from D0-brane dynamics}

In the previous section, we have seen that the matrix model proposed
in \cite{bmn} arises directly as a regularized version of the
light-cone supermembrane action in the maximally supersymmetric
pp-wave background of eleven-dimensional supergravity, just as
ordinary Matrix theory arises from the flat space supermembrane. 

In the usual case, an alternate approach to deriving the matrix model as a 
description of M-theory is by showing an equivalence between the DLCQ of 
M-theory with $N$ units of lightlike momentum \cite{susskind} and a limit of 
type 
IIA string theory with $N$ D0-branes in which the only remaining degrees of 
freedom are the lightest open strings between the D0-branes \cite{bfss, seiberg, 
sen}. In 
this section, we will try to obtain the matrix model of BMN directly from this 
D0-brane perspective.  

\subsection{Relation to type IIA string theory}

In the usual arguments \cite{sen,seiberg}, DLCQ M-theory with $N$ units of 
momentum is 
obtained (or defined) by taking a compactification of M-theory on a spacelike 
circle of radius $R_s$, performing a boost with parameter 
\be
\gamma = \sqrt {{R^2 \over 2R_s} + 1}
\ee
and taking the limit $R_s \to 0$ where the boost parameter becomes infinite 
leaving a lightlike circle of fixed radius R. The original frame, with spacelike 
circle of radius $R_s$, is equivalent to type IIA string theory with $N$ 
D0-branes in a limit $g_s \to 0$, $\alpha' \to 0$ in which only the low energy 
D0-brane degrees of freedom remain. In order that the energies of the 
corresponding 
D0-brane states (minus the diverging constant piece $N/R_s$) match with the 
light cone energies in the DLCQ description, we must also rescale all 
dimensionful quantities in the D0-brane theory by $m \to (R/R_s)m$
before taking the limit.

In the present case, we begin with a non-flat background for DLCQ M-theory, 
namely the pp-wave metric. Applying the same arguments, we should find that the 
matrix model appropriate to this background is that describing D0-branes in a 
type IIA background obtained by (un)boosting and compactifying the eleven-
dimensional pp-wave background. 

To determine the relevant IIA background in the present case (where the 
background fields are not necessarily weak), note that under a boost to the 
speed $v$ in $x^{\pm}$ plane, 
\[
x^+\longrightarrow \sqrt{{1-v\over 1+v}}x^+\ ,\ \ \ \ 
x^-\longrightarrow \sqrt{{1+v\over 1-v}}x^-\ . 
\]
Hence, the pp-wave metric is boost covariant, and the effect of the boost is 
simply to rescale $\mu$. In fact, up to a factor of $\sqrt{2}$, this rescaling 
is later removed when we rescale dimensionful parameters to make light-cone 
energies match with the energies of D0-brane states, so we will take both into 
account at once simply by replacing $\mu \to \mu/\sqrt{2}$. More details of the 
boost and rescaling may be found in \cite{tv0}, where this procedure was 
carried out in detail for an arbitrary weak background.

Taking into account the rescaling, the metric for M-theory in the ``unboosted'' 
frame, which we should compactify on a spacelike circle to obtain the desired 
type IIA metric, is 
\be\label{ppmetric}
ds^2 = -(1+{F^2 \over 4}) dt^2 + (1-{F^2 \over 4}) (dx^{10})^2 - {F^2
\over 2}  dt dx^{10} 
+ \sum_{i=1}^3 dx_i dx_i + \sum_{a=4}^9 dx_a dx_a\ , 
\ee
where $F^2$  is defined in (\ref{bg}) and the four form field is
\be
F_{123 (10)}=-F_{0123}={1\over 2} \mu \ .
\ee

Compactifying along $x^{10}$, the corresponding IIA background (in string 
frame) is
\bea\label{10met}
&{}&ds^2_{10}=g_{00} dt^2+ g_{AB} dx^A dx^B\ , \cr
&{}& g_{00}= - e^{-2\phi \over 3}, ~~~~ g_{AB}= \delta_{AB} e^{2 \phi \over 3}, 
\ 
\ A,B=1,\cdots , 9\, 
\eea
and
\bea 
e^{4\phi\over 3} = 1 - {1\over 4} F^2\ &,&\ \ \
C_{0} = {G_{0 10}\over G_{10,10}} = -{F^2 \over {4- F^2}} \\
 F_{0123} = -{\mu \over 2}\ &,& ~~~~H_{123} = {\mu \over 2} 
\label{elevtoten}
\eea 
where $C_0$ is the (time-component) of the RR-one form potential and 
$H_{123}$ and $F_{0123}$ are the NSNS and RR-three form field strengths, 
respectively.

Although the eleven dimensional pp-wave metric that we started with has 
zero scalar curvature (and its only nonvanishing component of Ricci scalar 
is $R_{++}={1\over 2}\mu^2$), the ten dimensional metric has a non-zero 
curvature. This can be understood noting that the effective 
compactification radius ($e^{2\phi\over 3}$) has a non-trivial 
$x$-dependence. 
By explicit calculation one can work out the scalar curvature of the 
metric (\ref{10met}):
\be
{\cal R}=- {\mu^2\over 8}{\Big (}1 - F^2/4 {\Big )}^{-{3\over 2}}
\left[ a - b~ \mu^2 \sum_{i=1}^3 x_ix_i - c~ \mu^2 \sum_{a=4}^9 x_a
x_a\right]\ ,
\ee
where $a,b,c$ are positive constants. 

We can trust the above SUGRA solution only when the curvature is 
small, i.e. when $F^2 \ll 4$. In fact, we see that at $F^2 = 4$, the background 
becomes singular, and is ill-defined beyond this point. From the eleven 
dimensional point of view, the reason for this bad behavior is that the $x^{10}$ 
direction becomes timelike beyond this point, so clearly, compactification does 
not make sense. Nevertheless, we may hope that in the region for which the type 
IIA background is well behaved, the action describing D0-branes in this 
background may be a valid description of the regime in which the DLCQ M-theory 
degrees of freedom stay within the weakly coupled region of the pp-wave 
background.

In this weakly curved limit, the IIA background fields can be expanded in powers 
of $F^2$. Up to the first order in $F^2$ we find
\bea\label{weakexp}
&& e^{4\phi \over 3} = 1-{F^2\over 4} \Rightarrow \phi \sim 
-{3\over 16}F^2 \cr 
&& C_0 = - {F^2 \over 4 - F^2} \sim - {F^2\over 4} \cr
&& g_{00} = - e^{-2\phi \over 3} \sim -1 - {1\over 8} F^2 \cr
&& g_{AB} = e^{2\phi \over 3} \delta_{AB} \sim \delta_{AB} - {1\over 8}F^2
\delta_{ij}\ .
\eea
We proceed in the next subsection to determine the action for D0-branes in this 
background.

\subsection{D0-brane dynamics on compactified reduced pp-waves}

We have seen that it is only possible to relate the eleven-dimensional pp-wave 
metric to a reliable type IIA supergravity background in the region where $F^2 
\ll 4$. In this weakly curved region, we may read off the action
describing 
D0-branes in the background (\ref{weakexp}) from the results of
\cite{tv0}, which 
provides the action for D0-branes at low energies and weak coupling in a general 
weak background.

In the notation of that paper, terms arising from the eleven-dimensional metric 
for our background are
\bea
S_2 & =& {1 \over 2} \partial_A \partial_B \phi I_{\phi}^{(AB)} + {1\over 4} 
\partial_A \partial_B h_{\mu \nu} I_h^{\mu \nu (AB)} + {1 \over 2} \partial_A 
\partial_B C_0 I_0^{0(AB)} 
\eea
where, for example $I_\phi^{(AB)}$ is the $AB$ moment of the current coupling 
to $\phi$. Using the expressions for the currents $I$ given in \cite{tv0} and 
the background fields given in (\ref{weakexp}), we obtain 
\[
S_2 = -{1\over 4} \partial_A \partial_B F^2 (T^{++(AB)} + {\rm higher \; 
orders} )\ ,
\] 
where 
\[
T^{++(AB)} = {1 \over R} \tr(X^A X^B)
\]
and the higher order terms vanish in the $\alpha' \to 0$ limit we are taking. 
Plugging in our expression for $F^2$, we obtain exactly the bosonic mass terms 
in the $BMN$ matrix model.

To obtain the terms arising from the eleven-dimensional three-form field, the 
required terms in the zero-brane action are
\bea
S_3 &=& \partial_A B_{\mu\nu} I_s^{\mu\nu (A)} + 
\partial_A  C_{\alpha\beta\gamma} I_2^{\alpha\beta\gamma(A)}
\eea 
To obtain the field strengths (\ref{elevtoten}), we may choose potentials
\be
B_{ij} = {\mu\over 6} \epsilon_{ijk}x_k, ~~
C_{0ij} = {\mu\over 6} \epsilon_{ijk}x_k
\ee
Using these expressions and the expressions given for the currents in 
\cite{tv0}, we find 
\[
S_3 = \mu \epsilon_{ijk} (J^{+ij(k)} + {\rm higher \; orders})
\]
where 
\[
J^{+ij(k)} = -{1 \over 6R} \tr( i [X^i ,X^j] X^k + {i \over 8} \Psi^\top 
\gamma^{ijk} \Psi ) 
\]
and again, the higher order terms vanish in the limit we are considering. Thus, 
the form-field background gives rise to both the fermion mass terms and the 
bosonic cubic interaction appearing in the BMN matrix model.

We have now seen that all terms in the BMN matrix model may be reproduced by 
considering D0-branes in an appropriate weak supergravity 
background.\footnote{These results could also be obtained directly using the 
general results for matrix theory in an arbitrary weak background \cite{tvm}.} 
From this point of view, it is remarkable that the model does not receive any 
corrections, since we have seen that the type IIA background becomes strongly 
curved and unreliable as we move far from the origin and in fact fails to make 
sense for $F^2 \ge 4$.

\section{Classical supersymmetric vacua}

In this section, we begin a direct study of the matrix model. We first write 
down the action in a convenient form and discuss the symmetries of the theory 
and their consequences for the spectrum. We then recall the supersymmetric vacua 
of the model and expand the matrix theory action about each of these solutions. 
We find that near each vacuum, the theory is described by a quadratic 
Hamiltonian with interactions suppressed by ${1 \over \mu}$.

\subsection{Action and symmetries}

It will be most convenient to write the action in a form that makes the 
$SO(3) \times SO(6)$ symmetries manifest, so we divide the scalars into $X^i$ 
with
$i = 1,2,3$ and $X^a$ with $4 \le a \le 9$, and write the fermions as $\psi_{I 
\alpha}$ where $I$ is a fundamental index of $SU(4) \sim SO(6)$ and $\alpha$ is 
a fundamental index of $SU(2) \sim SO(3)$. The relation between these fermions 
and the real 16 component spinor $\Psi$ is detailed in appendix A.

With these conventions, the matrix model describing the DLCQ of M-theory in the 
maximally supersymmetric pp-wave background is given by\footnote{Throughout this 
paper, we set $l_P = 1$, but we can restore $l_P$ in any formulae using the fact 
that $R$ and $1/\mu$ have dimensions of length.} 
\bea
{\cal L} &=&  \tr \left( {1 \over 2R} D_0 {X}^i D_0 {X}^i + 
{1 \over 2R} D_0 X^a D_0 X^a + i \psi^{\dagger I \alpha} D_0 \psi_{I \alpha} 
\right)\cr
 && \! \! \! \! \! \! \! \! \! \! + R \; \tr \left(
- {1 \over 2} \left({\mu \over 3R}\right)^2 (X^i)^2  
- {1 \over 2} \left({\mu \over 6R} \right)^2 (X^a)^2  - {\mu \over 4R} 
\psi^{\dagger I \alpha} \psi_{I \alpha} - {i \mu \over 3R} \epsilon_{ijk} X^i 
X^j X^k  
\right. \cr
&& \qquad 
- \psi^{\dagger I \alpha} \sigma^i_\alpha {}^\beta [X^i, \psi_{I \beta}] 
+ {1 \over 2} \epsilon_{\alpha \beta} \psi^{\dagger \alpha I} {\sf g}^a_{IJ} 
[X^a, \psi^{\dagger \beta J}] - {1 \over 2} \epsilon^{\alpha \beta} 
\psi_{\alpha I} ({\sf g}^{a \dagger})^{IJ} [X^a, \psi_{\alpha J}] \cr
&& \left. \qquad +  { 1 \over 4} [X^i , X^j]^2 
+ {1 \over 4} [X^a, X^b]^2 + { 1 \over 2} [ X^i, X^a]^2 \right)
\eea
Here, $\sigma^i$ are the usual Pauli matrices and ${\sf g}^a_{IJ}$ relate the 
antisymmetric product of two $SU(4)$ fundamentals to an $SO(6)$ fundamental.

The new terms relative to the flat space matrix model appear in the second line. 
These include mass terms for each of the matrix theory variables, as well as a 
cubic interaction which favors matrix configurations carrying a dipole moment of 
membrane charge. 

As for the case of flat space, the $U(1)$ part of the theory (describing the 
center of mass degrees of freedom) decouples from the $SU(N)$ part. In this 
case, the $U(1)$ sector is described by a harmonic oscillator Hamiltonian 
with 
bosonic oscillators in the $SO(6)$ directions of mass ${\mu \over 6}$ and in 
the $SO(3)$ directions of mass ${\mu \over 3}$, as well as 8 fermionic 
oscillators of mass ${\mu \over 4}$. Thus, unlike the flat space case, the 
different polarization states have different masses.

In appendix B, we review the full symmetry algebra of the matrix model and 
provide explicit expressions for the generators in terms of the matrix theory 
variables. The bosonic generators include the Hamiltonian, rotation generators 
in the $SO(3)$ and $SO(6)$ directions, the central light-cone momentum $P^+ = {N 
\over R}$, and generators $a_i$ and $a_a$ which are exactly the harmonic 
oscillator creation operators from the $U(1)$ part of the theory. In addition, 
we have 32 fermionic symmetries.  Sixteen of the generators are the 8 
complex fermionic creation operators from the $U(1)$ part of the theory, 
\[
q_{I \alpha} = {1 \over \sqrt{R}} \tr(\psi_{I \alpha}) 
\]
which generate the overall polarization states. The other 16 supersymmetries are 
generated by 
\beas
Q_{I \alpha} &=& \sqrt{R} \tr \left( (\Pi^a - i {\mu \over 6R} X^a)
{\sf g}^a_{IJ} 
\epsilon_{\alpha \beta} \psi^{\dagger J \beta} - (\Pi^i + i {\mu \over 3R} X^i)
 \sigma^i_\alpha {}^\beta \psi_{I \beta} \right.\\
& & \left. + {1 \over 2} [X^i, X^j] \epsilon^{ijk} \sigma^k_\alpha {}^\beta 
\psi_{I \beta} - {i \over 2} [X^a, X^b] ({\sf g}^{ab})_I^J \psi_{J \alpha} +i 
[X^i, X^a] \sigma^i {\sf g}^a_{IJ} \epsilon_{\alpha \beta} \psi^{\dagger I 
\beta} \right)
\eeas
which include new terms linear in $X$ relative to the flat space model. 

The supersymmetry algebra has a number of unusual features compared with the 
flat space case. First, the supercharges do not commute with the Hamiltonian but 
rather obey commutation relations
\beas
\left[H, q_{I \alpha} \right] &=& -{\mu \over 4} q_{I \alpha}\\ 
\left[H, Q_{I \alpha} \right] &=& {\mu \over 12} Q_{I \alpha}
\eeas
Thus, different members of a multiplet of states generated by acting with 
supercharges will have different energies, although the energy differences will 
still be exactly determined by the symmetry algebra. In addition, we now have 
rotation generators appearing on the right hand side of the $\{ Q, Q \}$ 
commutation relation along with the Hamiltonian, 
\[
\{ Q^{\dagger I \alpha}, Q_{J \beta} \} = 2 \delta^I_J \delta^\alpha_\beta H   
- {\mu \over 3} \epsilon^{ijk} \sigma^k_{\beta} {}^\alpha \delta^I_J
M^{ij} - {i \mu \over 6}  \delta^\alpha_\beta ({\sf g}^{ab})_J {}^I
M^{ab} \; .
\]
Note that we still have $\{ Q^{\dagger I \alpha}, Q_{I \alpha} \} \propto H$, so 
that a state is fully supersymmetric if and only if it has zero
energy. On the other hand, unlike the ordinary matrix model, we now
have the possibility of BPS states with non-zero energies carrying
some angular momentum. We will see that 1/4, 3/16, 1/8, and 1/16 BPS states are 
possible in addition to the ground states which are 1/2 BPS (preserving 16 
supercharges).
In this case, the rotation generators are not central charges since they do not 
commute with the supersymmetry generators. As a result, the representation 
theory of this algebra is quite different from the usual case, and in 
particular the notion of BPS multiplets is altered. We will
discuss the BPS states in more detail in section 5.5.

\subsection{Classical supersymmetric solutions}

We now recall the classical supersymmetric vacua of the matrix model
given in \cite{bmn}. To begin, we note that the bosonic potential may be written 
as a sum of 
positive definite terms,
\[
V = {R \over 2}\ {\rm Tr}\left[ \left({\mu \over 3R} X^i + i \epsilon^{ijk} 
X^j X^k\right)^2 + { 1\over 2 } ( i [X^a, X^b])^2 + (i [X^a, X^i])^2 + 
\left({\mu \over 6R}\right)^2 (X^a)^2 \right].
\]     
For supersymmetric solutions, each of the four terms must vanish independently. 
It is clear that for the last term to vanish, we must take $X^a$ = 0, while 
for the first term to vanish we must have
\[
X^i = {\mu \over 3R} J^i
\]
where $J^i$ form a representation of the $SU(2)$ algebra, 
\[
[J^i, J^j] = i \epsilon^{ijk} J^k
\]
Thus, the classical supersymmetric solutions are labeled by the $N$ dimensional 
representations of $SU(2)$, and the number is given by the number of partitions 
of $N$ into sums of positive integers, as noted in \cite{bmn}. 

Geometrically, a solution labeled by a partition $\{ N_1, \dots , N_k \}$ 
corresponds to a set of membrane fuzzy spheres with radii $r_i \sim {N_i \mu 
\over 6R}$. The spheres are stabilized through a competition between the 
quadratic and quartic terms in the potential, which make the spheres want to 
contract, and the cubic term which makes the spheres want to expand. The net 
effect is that any membrane in the pp-wave background expands to a preferred 
radius proportional to its light-cone momentum. Unlike the similar situation 
studied by Myers \cite{myers}, the various solutions all have zero energy and 
are thus 
classically stable. 

It is particularly interesting that the classical vacuum states each correspond 
to a definite number of membranes. Thus, at low energies, one of the difficult 
features of the usual matrix model, distinguishing between single and multi-
membrane states, is absent.

\subsection{Expansion about the classical vacua}

We now consider expanding the action about its classical vacua, that is taking
\be
\label{expand}
X^i = {\mu \over 3R} J^i + Y^i 
\ee
for a given choice of $J$.

It is convenient to rescale variables to remove all dependence on parameters 
from the quadratic action, so we take 
\be\label{scale}
Y^i \to \sqrt{R \over \mu} Y^i, \; X^a \to \sqrt{R \over \mu} X^a, 
\; t \to {1 \over \mu} t  
\ee
Note that energies will be measured in units of $\mu$ from now on; we will 
restore this factor in the important formulae.

Plugging (\ref{expand}) into the action and performing the rescalings, we 
find that the resulting action (in the $A_0=0$ gauge) takes the form
\[
S = S_2^Y + S_2^X + S_2^\psi + S_3 + S_4
\]
where 
\bea
S_2^Y &=&  \tr \left( {1 \over 2} \dot{Y}^i \dot{Y}^i - { 1 \over 2} 
({1 \over 3})^2(Y^i + i \epsilon^{ijk} [J^j, Y^k])^2 \right) \cr
S_2^X &=& \tr \left( {1 \over 2} (\dot{X}^a)^2 - {1 \over 2} ({ 1\over 3})^2 ( 
{1 \over 4} (X^a)^2 - [J^i, X^a]^2 )  \right) \cr
S_2^\psi &=& \tr \left( i \psi^\dagger \dot{\psi} - {1 \over 4} 
\psi^\dagger \psi - {1 \over 3} \sigma^i \psi^\dagger [J^i, \psi_\beta] \right)
\eea
are quadratic actions for $X$, $Y$, and $\psi$, and
\bea
S_3 &=& \left( {R \over \mu} \right)^{3\over 2}  \tr \left( {1 \over 3} 
[J^i, X^a][Y^i, X^a] + 
{1 \over 3} [J^i, Y^j][Y^i, Y^j] - { i \over 3} \epsilon^{ijk} Y^i Y^j Y^k 
\right. 
\cr
 & &  
\qquad \qquad  \left. - \psi^{\dagger} \sigma^i [Y^i, \psi] 
+ {1 \over 2} \epsilon \psi^{\dagger} {\sf g}^a 
[X^a, \psi^{\dagger}] - {1 \over 2} \epsilon \psi ({\sf g}^\dagger)^{a} [X^a, 
\psi] \right) \cr   
S_4 &=& \left({R \over \mu} \right)^3 \tr \left( { 1 \over 4} [Y^i , Y^j]^2 
+ {1 \over 4} [X^a, X^b]^2 + { 1 \over 2} [Y^i , X^a]^2 \right)
\eea
are the interaction terms. 

We see that for large ${\mu \over R}$, the interaction terms are suppressed by 
powers of ${R \over \mu}$ relative to the quadratic action, so we should be able 
to study the low energy excitations about these classical vacua using 
perturbation theory in ${R \over \mu}$. In particular in the ${\mu\over R} 
\to \infty$ limit, the theory divides into superselection sectors labeled by 
$N$-dimensional representations of $SU(2)$ and each section will be described by 
a set of bosonic and fermionic harmonic oscillators. Our next step will be to 
solve the quadratic theory explicitly in this limit before considering the 
effects of interactions in section 6.

\section{Exact spectrum in the large $\mu$ limit}

In the limit of large $\mu$ for fixed $N$ and $R$, the various classical vacua 
become separated by infinite energy barriers and may be treated as 
superselection sectors, each described by a quadratic Hamiltonian.
As a first step towards understanding the theory, we would now 
like to explicitly diagonalize the quadratic actions in order to find the 
complete spectrum in the limit of large $\mu$. 

\subsection{Expansion in terms of matrix spherical harmonics}

To proceed, we note that many of the terms in the quadratic action are written 
in terms of commutators with the $J^i$ matrices,
\[
[J^i, A]
\]
It will be convenient then to expand our fluctuation matrices in a basis which 
behaves simply under this commutator action. To do this, note that the set of $N 
\times N$ matrices forms an $N^2$-dimensional reducible representation of 
$SU(2)$, with the $SU(2)$ generators acting as above. It is not hard to see that 
this representation decomposes into irreducible representations as
\[
N^2 = 1 \oplus 3 \oplus \cdots \oplus (2N-1) \; ,
\]
that is, integer spins up to $j = N-1$. We may therefore expand an 
$N \times N$ matrix as
\[
A = \sum_{j=0}^{N-1} \sum_{m=-j}^j a_{jm} Y_{jm} 
\]
where $Y_{jm}$ transform in the irreducible spin $j$ representation. For 
Hermitian matrices, we must have the reality condition
\be
\label{real}
a^*_{jm} = (-1)^m a_{j \; -m} \; .
\ee

Explicit expressions for these matrix spherical harmonics may be obtained by 
writing the usual spherical harmonics as symmetric polynomials in $x_i$ with 
$x_1^2 + x_2^2 + x_3^2 = 1$ and replacing $x$'s with $J$s (up to normalization). 
In this way, we find 
that the spin $j$ representation may be written as products of $j$ $J^i$ 
matrices whose indices are contracted with symmetric, traceless tensors. A 
detailed discussion may be found in \cite{hoppe}.

For example 
\bea
Y_{00} = 1 && \qquad  Y_{1 \;-1} =  
\sqrt{{6 \over (N^2-1)}} J^-  \cr
 Y_{10} = \sqrt{{12 \over (N^2-1)}} J^3 && \qquad
Y_{11} = -\sqrt{{6 \over (N^2-1)}} J^+, \dots 
\eea
where $J^{\pm} = J^1 \pm i J^2$. In general, we will choose normalizations so 
that
\be\label{Ynorm}
\tr( Y^\dagger_{j'm'} Y_{jm}) = N \delta_{j' \; j} \delta_{m' \; m} \; .
\ee
With these definitions, we may use the usual properties of angular momentum 
generators, 
\bea
[J_3, Y_{jm}] = m Y_{jm} , \qquad && [J^+, Y_{jm}] = \sqrt{(j-m)(j+1+m)} Y_{j \; 
m+1} \cr
  [J^i, [J^i, Y_{jm}]] = j(j+1) Y_{jm} && [J^-, Y_{jm}]  = \sqrt{(j+m)(j+1-m)} 
Y_{j \; m-1}
\eea
We now proceed to diagonalize each of the quadratic actions above.

\subsection{Spectrum of the irreducible vacuum}

We begin by considering the single-membrane vacuum corresponding to the $N$ 
dimensional irreducible representation of SU(2). 

\subsubsection*{Bosons in the SO(6) directions}

From the action above, we see that the quadratic potential for $X^a$ may be 
written as
\[
V^X = {1 \over 2} { 1\over 3^2} \tr {\Big (}X^a {\Big (} {1 \over 4} X^a  + 
[J^i, [J^i, 
X^a]]{\Big )}{\Big )} 
\]
In this form, it is obvious that the mass matrix is diagonal in the basis of 
irreducible representations of $SU(2)$, 
\[
X^a  = {1 \over \sqrt{N}} x^a_{j m} Y_{jm}
\]
and the eigenvalues are 
\[
M^2 = {\Big (}{1 \over 3} {\Big (}j + { 1 \over 2}{\Big )}{\Big )}^2
\]
each with degeneracy 6. The quadratic Lagrangian becomes
\[
{\cal L} = \sum_{j=0}^{N-1} \; {1 \over 2} |\dot{x}^a_{jm}|^2 - 
{1 \over 2} {\Big (}{1 
\over 3} 
{\Big (}j + {1 \over 2} {\Big )}{\Big )}^2 |x^a_{jm}|^2 
\]
where $j$ runs from 0 to $N-1$ and $m$ runs from $-j$ to $j$, as usual.

\subsubsection*{Bosons in the SO(3) directions}

For the $Y^i$ oscillators, we may rewrite the quadratic potential as
\[
V^Y = {1 \over 2} ({ 1\over 3})^2 (Y^i + i \epsilon^{ijk} [J^j, Y^k])^2  
\]
Thus, if we solve the eigenvalue problem 
\[
Y^i + i \epsilon^{ijk} [J^j, Y^k] = \lambda Y^i 
\]
the masses will be given as $M^2 = ({1 \over 3} \lambda)^2$.\footnote{Here, the 
matrix $Y^i$ transforms in the tensor product of the spin 1 representation of 
$SU(2)$ (from the explicit vector index) with the sum of the spin $j$ 
representations. The eigenvectors are precisely the irreducible 
representations, 
corresponding to vector spherical harmonics.}

To find the eigenvalues and eigenvectors, we write 
\[
Y^i = y^i_{jm} Y_{jm}\ . 
\]
Making this substitution and writing out the three equations corresponding to 
$i = +, -, 3$, we find a set of equations
\bea
&&(\lambda + m) y^+_{j \; m+1} = \sqrt{j(j+1) - m(m+1)} y^3_{j m}  \cr
&&(\lambda - m) y^-_{j \; m-1} = - \sqrt{j(j+1) - m(m-1)} y^3_{j m} \\
&&(\lambda - 1) y^3_{j \; m}   = {1 \over 2} \sqrt{j(j+1) - m(m+1)} 
y^+_{j \; m+1} - {1 \over 2} \sqrt{j(j+1) - m(m-1)} y^-_{j m-1} \nonumber
\eea
Note that we have a separate set of equations for each value of $m$ between 
$m = -j-1$ and $m = j+1$.\footnote{For $m=\pm j$, one of the equations vanishes 
while for $m= \pm(j+1)$, two of the equations vanish.} Replacing $y^+$ and $y^-$ 
in the third equation using the first two equations, we find an equation for the 
eigenvalues, 
\[
\lambda(\lambda + j) (\lambda - (j + 1)) = 0
\]
so we may have $\lambda = 0$,  $\lambda = -j$ or $\lambda = j+1$. Using the 
first two equations, we can read off the eigenstates for these three 
eigenvalues. For $\lambda = 0$, the eigenstates are
\[
Y^i = [J^i, Y_{jm}] \qquad 1 \le j \le N-1, \; -j \le m \le j
\]
for a total of $N^2-1$ zero modes. These correspond to fluctuations in the
 directions of the gauge orbit and not to physical zero-modes. To see this, 
note that under an infinitesimal gauge transformation, 
\[
{1 \over 3} J^i \to {1 \over 3} J^i + {1 \over 3} i [J^i, \Lambda]
\]
so that any fluctuation of the form 
\[
 Y^i  = [J^i, \Lambda ]
\]
corresponds to a gauge transformation. Since $[J^i , Y_{00}] = 0$, we
have $N^2-1$ instead of $N^2$ gauge orbit directions, as we found above.

For the $\lambda = -j$ eigenstates, corresponding to oscillators with 
$M = {j \over 3}$, we find
\bea
Y^+ &=& - {\alpha_{j-1 m} \over \sqrt{N}} \sqrt{(j+m)(j+1+m) \over j(2j+1)}  
Y_{j \; m+1} \cr
Y^- &=&   {\alpha_{j-1 m} \over \sqrt{N}}  \sqrt{(j-m)(j+1-m) \over j(2j+1)}  
Y_{j \; m-1} \cr
Y^3 &=&  { \alpha_{j-1 m} \over \sqrt{N}} \sqrt{(j+m)(j-m) \over j(2j+1)} Y_{jm} 
\eea  
where $0 < j < N$ and $ -j < m < j $ and we have the reality condition 
$\alpha^*_{jm} = (-1)^m \alpha_{j \; -m}$.

For the $\lambda = (j+1)$ eigenstates, corresponding to oscillators with 
$M = {j+1 \over 3}$, we find
\bea
Y^+ &=&  {\beta_{j+1 \; m}\over \sqrt{N}}  \sqrt{(j-m)(j+1-m) \over (j+1)(2j+1)} 
Y_{ j \; m+1}
\cr
Y^- &=&  - {\beta_{j+1 \; m}\over \sqrt{N}}  \sqrt{(j+m)(j+1+m) \over 
(j+1)(2j+1)} Y_{ j \; m-1 } \cr
Y^3 &=&  {\beta_{j+1 \; m}\over \sqrt{N}} \sqrt{(j+1+m)(j+1-m) \over 
(j+1)(2j+1)} Y_{j \; m } 
\eea  
where $0 \le j < N$ and $-j-1 \le m \le j+1$ and we have the reality condition
$\beta^*_{jm} = (-1)^m\beta_{j -m}$.

In terms of $\alpha$ and $\beta$, the quadratic Lagrangian becomes
\beas
{\cal L}^Y = && \sum_{j=0}^{N-2} \; {1 \over 2} |\dot{\alpha}_{jm}|^2 - {1 \over 
2} ({j + 1 \over 3})^2 |\alpha_{jm}|^2 \\
+&& \sum_{j=1}^{N} \;  {1 \over 2} |\dot{\beta}_{jm}|^2 - {1 \over 2} ({j 
\over 3})^2 |\beta_{jm}|^2
\eeas

\subsubsection*{Fermions}

The quadratic fermion potential may be written as 
\be
\label{ferm}
V^\psi = \psi^{\dagger \alpha} ( {1 \over 4} \psi_\alpha + {1 \over 3} 
\sigma^i_\alpha {}^{\beta} [J^i, \psi_\beta]) 
\ee
where we have suppressed $SU(4)$ indices. Thus, we would like to find the 
eigenstates of the equation\footnote{In this case, we are 
combining the spin $j$ representations with the spin ${1 \over 2}$ 
representation, and again, the eigenstates will simply be the irreducible 
representations.}
\[
\sigma^i_\alpha {}^\beta [J^i, \psi_\beta] = \lambda \psi_\beta 
\]
in terms of which the fermion masses will be
\[
m = |{1 \over 4} + {1 \over 3} \lambda|
\]
Making the expansion
\[
\psi_\alpha = \psi_\alpha^{jm} Y_{jm}
\]
and inserting into the eigenvalue equation, we find two equations
\beas
(\lambda - m) \psi_+^{jm} &=& \sqrt{(j-m)(j+1+m)} \psi_-^{j \; m+1}\\
(\lambda + m + 1) \psi_-^{j \; m+1} &=& \sqrt{(j-m)(j+1+m)} \psi_+^{jm}
\eeas
where we have labeled the $SU(2)$ index values by $+$ and $-$ to denote the 
$\sigma^3$ eigenvalue. Combining these yield the equation
\[
(\lambda + j + 1)(\lambda - j) = 0
\]
so the masses are given by $M={1 \over 3} (j+ {3 \over 4})$ and $M={1 \over 
3}(j+ {1 \over 4})$. Explicitly, the $M={1 \over 3} (j+ {3 \over 4})$  
eigenstates are
\beas
\psi_+ &=& {\eta^{j+ {1 \over 2} \; m + {1 \over 2}} \over \sqrt{N}} \sqrt{j+1+m 
\over 2j+1} Y_{jm}\\
\psi_- &=& {\eta^{j+ {1 \over 2} \; m + {1 \over 2}} \over \sqrt{N}} \sqrt{j-m 
\over 2j+1} Y_{j \; m+1}
\eeas
where we may have $0 \le j \le N-1$ and $-j-1 \le m \le j$. The $M={1 \over 
3}(j+ {1 \over 4})$ eigenstates are 
\beas
\psi_+ &=& -{(\chi^\dagger)^{j - {1 \over 2} \; m + {1 \over 2} } \over 
\sqrt{N}} 
\sqrt{j-m \over 2j+1} Y_{j \; m}\\
\psi_- &=& {(\chi^\dagger)^{j - {1 \over 2} \; m + {1 \over 2} } 
\over \sqrt{N}} 
\sqrt{j+1+m \over 2j+1} Y_{j \; m+1}\\
\eeas
where $1 \le j \le N-1$ and $-j \le m \le j-1$.

In terms of the oscillators $\eta$ and $\chi$, the quadratic Lagrangian for 
the fermions becomes
\beas
{\cal L}^\psi = &&\sum_{j={1 \over 2}}^{N -{3 \over 2}} \; i \chi^\dagger_{j 
 m} D_0 \chi_{jm} - {1 \over 3}(j + {3 \over 4}) \chi^\dagger_{j m} 
\chi_{jm} \\
+&& \sum_{j={1 \over 2}}^{N-{1 \over 2}} \; i \eta^\dagger_{jm} D_0 
\eta_{jm} - {1 \over 3}(j + {1 \over 4}) \eta^\dagger_{ j m} \eta_{j m}
\eeas
Here, $\eta$ and $\chi$ carry fundamental and antifundamental $SU(4)$ indices 
respectively, which we have suppressed.

\subsubsection*{Summary of oscillator masses for the irreducible vacuum}

The spectrum of oscillators about the single membrane vacuum is summarized in 
table 1, with the masses given in units of $\mu$.

\begin{table}[ht]
\begin{center}
\begin{tabular}{|c||c|c|c|c|c|}
\hline
{\rm Type} & {\rm Label} & {\rm Mass} & {\rm Spins} & $SO(6) \times 
SO(3)$ {\rm Rep} & {\rm Degeneracy} 
\\ \hline\hline
$S0(6)$ & $x^a_{jm}$ & ${1 \over 6} + {j \over 3}$ & $0 \le j \le N-1$ 
& $(6, 2j+1)$ & $6(2j+1)$ 
\\ \hline
$S0(3)$ & $\alpha_{jm}$ & ${1 \over 3} + {j \over 3}$ & $0 \le j \le N-2$ 
& $(1, 2j+1)$ & $(2j+1)$
 \\ 
  & $\beta_{jm}$ & ${j \over 3}$ & $1\le j \le N$ & $(1, 2j+1)$ & $(2j+1)$ 
\\ \hline
{\rm Fermions} & $\chi^I_{jm}$ & ${1 \over 4} + {j \over 3}$ & 
${1 \over 2} \le j \le N - {3 \over 2}$ & $(\bar{4}, 2j+1)$ & $4(2j+1)$ 
\\ 
   & $\eta_{I \; jm}$ & ${1 \over 12} + {j \over 3}$ & ${1 \over 2} \le j \le 
N - {1 \over 2}$ & $(4, 2j+1)$ & $4(2j+1)$ \\
\hline
\end{tabular} \caption{Oscillator masses for the irreducible vacuum}
\end{center}
\end{table}
It is easy to check that the sum of the boson masses and the sum of the fermion 
masses are both ${2 \over 9} N (8N^2 + 1)$, as required for the
vanishing of the vacuum 
state zero point energy. 

A nice feature of the spectrum we have calculated is that the oscillator masses 
are completely independent of $N$. As one might hope from the discretized 
membrane picture, $N$ acts exactly as a short distance cutoff, providing an 
upper bound on the allowed values of the spin $j$ (related to the wavelength of 
modes on the sphere). In particular, the spectrum has a well defined limit for 
large $N$, which should reproduce the spectrum of the quadratic part of the 
continuum supermembrane field theory for this background.\footnote{It is 
important to note that for any finite $\mu$, there will be some energy above 
which the perturbation theory breaks down and the spectrum can no longer be 
trusted. We discuss this further below.}

\subsection{Oscillator masses in a general reducible representation}

A general classical vacuum state of the theory corresponds to an $N$ dimensional 
reducible representation of $SU(2)$, described by a block-diagonal matrix
\[
X^i = \left( \ba{ccc} J^i_1 & & \\ & \ddots & \\ & & J^i_K \ea \right)
\]
where $J^i_l$ are generators of the $N_l$-dimensional irreducible representation 
of $SU(2)$ and $N_1 + \dots + N_K = N$.

To deal with these states, it is convenient to break up our matrices into blocks 
corresponding to the individual irreducible representations, so for example
\[
X^i_{kl} = \delta_{kl} J^i_l
\]
where $1 \le k,l \le K$ are indices labeling the blocks.
Then the quadratic actions split up into a sum of terms which are of same form 
as for the irreducible representation above,
\beas
V^X &=& {1 \over 2} ({ 1\over 3})^2 \tr((X_{kl}^a)^\dagger ( {1 \over 4}X^a_{kl}  
+ J^i 
\circ (J^i \circ X_{kl}^a))) \\
V^Y &=& {1 \over 2} |({1 \over 3})^2 (Y^i_{kl} + i \epsilon^{ijk} J^j \circ 
Y_{kl}^k)|^2  \\
V^\psi &=& \psi_{kl}^{\dagger \alpha} ( {1 \over 4} \psi_\alpha + {1 \over 3} 
\sigma^i_\alpha {}^{\beta} J^i \circ \psi_{kl \; \beta}) 
\eeas
but we now define
\[
\label{act}
J^i \circ B_{kl} = J^i_k B_{kl} - B_{kl} J^i_l \; .
\]
Thus, whereas the matrices in the irreducible case were in the tensor 
product of 
two $N$ dimensional representations of $SU(2)$ and split up into the 
representations $1 \oplus 3 \oplus \cdots \oplus 2N-1$, a general rectangular 
block $B_{kl}$ lies in the tensor product of the $N_k$ and $N_l$ dimensional 
representation and may be decomposed into irreducible representations with spins
${1 \over 2} |N_k - N_l| \le j \le {1 \over 2}(N_k + N_l) - 1$. Thus, for 
example, we may write
\[
X_{kl}^a = \sum (x_{kl}^a)_{jm} Y^{N_k N_l}_{jm} 
\]  
where $Y$'s are $N_k \times N_l$ matrices which are eigenstates of $J$ and $J_3$ 
acting as in (\ref{act}). Using this decomposition, the diagonalization proceeds 
exactly as above.

\begin{table}[ht]
\begin{center}
\begin{tabular}{|c||c|c|c|c|c|}
\hline
{\rm Type} & {\rm Label} & {\rm Mass} & {\rm Spins} & $SO(6) \times SO(3)$
\\ \hline\hline 
$S0(6)$ & $(x_{kl}^a)_{jm}$ & ${1 \over 6} + {j \over 3}$ & 
${1 \over 2}|N_k - N_l| \le j \le {1 \over 2} (N_k + N_l) - 1$ & $(6,2j+1)$ 
\\ \hline 
$S0(3)$ & $\alpha_{kl}^{jm}$ & ${1 \over 3} + {j \over 3}$ & ${1 \over 
2}|N_k - N_l| - 1 \le j \le {1 \over 2} (N_k + N_l) - 2$ & $(1,2j+1)$ 
\\ 
 & $\beta_{kl}^{jm}$ & ${j \over 3}$ & ${1 \over 2}|N_k - N_l| +1 \le j 
\le {1 \over 2} (N_k + N_l)$ & $(1,2j+1)$  
\\ \hline
{\rm Fermions} & $\chi^{I \;jm}_{kl}$ & ${1 \over 4} + {j \over 3}$ & 
${1 \over 2}|N_k - N_l| - {1 \over 2} \le j \le {1 \over 2} (N_k + N_l) - 
{3 \over 2}$ 
& $(\bar{4},2j+1)$ \\  
  & $\eta_{I \; kl}^{jm}$ & ${1 \over 12} + {j \over 3}$ & ${1 \over 2}|N_k - 
N_l| 
+ {1 \over 2} \le j \le {1 \over 2} (N_k + N_l) - {1 \over 2}$ & $(4,2j+1)$ 
\\
\hline
\end{tabular} \caption{Oscillator masses for reducible vacua}
\end{center}
\end{table}

The spectrum of oscillators about a general vacuum are given in table 2, again 
with masses in units of $\mu$. As before, we will have zero mass
oscillators 
from the $Y^i$ matrices 
corresponding to each of the non-trivial gauge orbit directions, and these will 
take the form
\[
Y^i = [J^i, \Lambda]
\]
for arbitrary $\Lambda$. The number of nontrivial gauge orbit directions will be 
$N^2$ minus the dimension of the space of matrices that commute with $J^i$, or
\[
N^2 - \sum M^2_i
\]
where $M_i$ is the multiplicity of the $i$'th irreducible representation. 

\subsection{Physical states}

We have now computed the spectrum of oscillators about an arbitrary vacuum of 
the theory. In the $\mu = \infty$ limit, one can view these classical 
solutions as superselection sectors, and the physical spectrum in each sector is 
obtained by acting on a Fock space vacuum with arbitrary numbers of creation 
operators for the oscillators found above. The only complication is that we must 
ensure that the Gauss law constraints, arising from the equation of motion for 
$A_0$, are satisfied. These $N^2-1$ conditions read
\[
i[X^i, \Pi^i] + i[X^a, \Pi^a] + 2 \psi_{I \alpha} \psi^{\dagger I \alpha} = 0
\]
The corresponding operator is ordered so that it becomes the generator of 
$SU(N)$ transformations, and the quantum constraint requires that this operator 
annihilates physical states. In other words, physical states must be $SU(N)$ 
invariant.

Expanding the constraint about $X^i = {\mu \over 3R} J^i$, we find 
\[
\left({\mu \over 3R} i [J^i, \Pi^i] +  i[Y^i, \Pi^i] + i[X^a, \Pi^a] + 2 
\psi_{I 
\alpha} \psi^{\dagger I \alpha}\right) |\psi \rangle = 0
\]
In the limit $\mu \to \infty$, it is useful to split these $N^2-1$ constraints 
into two groups depending on whether or not the first term vanishes. Recalling 
that normalized oscillators in the $SO(3)$ directions satisfy an eigenvalue 
equation
\[
\Pi^i + i \epsilon^{ijk} [J^j, \Pi^k] = \lambda \Pi^i
\]
with $M^2 = \lambda^2/3$, and commuting this equation with $J^i$, we find
\[
\lambda [J^i, \Pi^i] = 0 \; .
\]
Hence, the first term in the Gauss law constraint is non-zero precisely for the 
momenta associated with zero-mode ($\lambda = 0$) oscillators $Y^i = [J^i, 
\Lambda]$ 
corresponding to gauge orbit directions. For these $N^2 - M_i^2$ constraints 
(where $M_i$ is the multiplicity of the $i$th irreducible $SU(2)$ 
representation), we can ignore all but the first term in the large $\mu$ limit, 
and the condition is simply that the wavefunction does not depend on the 
coordinates corresponding to gauge orbit directions (i.e. that the zero-modes 
are non-physical). 

The remaining number of constraints, $\sum M_i^2 - 1$ is exactly the number of 
generators in the subgroup of $SU(N)$ under which the vacuum state is invariant. 
These constraints demand that physical states are invariant under this subgroup. 
In a vacuum with $M_i$ copies of the $N_i$ dimensional irreducible 
representation, the unbroken symmetry group will be $U(M_1) \times \cdots \times 
U(M_n)$ (including the diagonal U(1) whose Gauss law constraint is trivial). 
From table 2, we see that the oscillator spectrum corresponding to a given block 
depends only on the dimension of the two representations defining the block, so 
for each pair of representations $N_i,N_j$ we will have $M_i M_j$ identical 
copies of the same spectrum, which form a matrix in the bifundamental 
representation of $U(M_i) \times U(M_j)$ (or in the adjoint of $U(M_i)$ for the 
case $N_i = N_j$). The Gauss law constraint is then satisfied simply by forming 
$U(M_1) \times \cdots \times U(M_N)$ invariants out of the matrix creation 
operators. Thus, a general physical state will be obtained by acting with an 
arbitrary number of traces of products of these matrix creation operators on the 
fock space vacuum.

We have now specified the complete spectrum of physical states of the matrix 
model in the $\mu \to \infty$ limit.

\subsection{BPS States}

We have argued in section 4.1 that the unusual supersymmetry algebra
of the matrix model allows the possibility of states carrying angular
momentum which preserve various fractions of the 16 supersymmetries
generated by $Q$. Now that we have the complete spectrum of the
theory in the $\mu = \infty$ limit, it will be possible to explicitly
identify all BPS states.

To begin, it is convenient to write the anticommutation relation for
the $Q$s in the original $SO(9)$ notation,
\be
\label{QQ}
\{ Q_\alpha, Q_\beta \} = 2 \delta_{\alpha \beta} H - { \mu \over 3} 
(\gamma^{123} \gamma^{ij})_{\alpha \beta} M^{ij} + { \mu \over 6} (\gamma^{123} 
\gamma^{ab})_{\alpha \beta} M^{ab}
\ee
We may choose to label states by their eigenvalues for
a set of Cartan generators of the $SO(3) \times SO(6)$ algebra, which
we take to be
$M^{12}$, $M^{45}$, $M^{67}$, and $M^{89}$. The $\gamma$ matrices coupling to 
each of these generators all commute with one another and may be represented by
\bea
\gamma^3 &=& \sigma^3 \times 1 \times 1 \times 1 \qquad \gamma^{12345} = 1 
\times \sigma^3 \times 1 \times 1 \cr
 \gamma^{12367} &=& 1 \times 1 \times \sigma^3 \times 1  \qquad \gamma^{12389} = 
- 1 \times \sigma^3 \times \sigma^3 \times 1 \; . \label{gam}
\eea
where the last definition is implied by 
\[
\gamma^{12345} \gamma^{12367} \gamma^{12389} = - \gamma^{123456789} = -1 \; .
\]
Any non-trivial BPS state must have non-zero angular momentum
and therefore will lie in some non-trivial representation of $SO(3)
\times SO(6)$. For this $SO(3) \times SO(6)$ multiplet, there will be a unique 
highest weight state $|\psi \rangle$, and for this state, we will have $\langle 
M^{AB} \rangle = 0$ for all of the non-Cartan generators of the
algebra. Using the representation (\ref{gam}), the 
expectation value of (\ref{QQ}) on this state is then
\[
\langle \psi | \{ Q_\alpha , Q_\beta \} | \psi \rangle = 2 \delta_{\alpha \beta} 
\Delta_\beta  
\]
where the sixteen diagonal elements $\Delta_\alpha$ are given by two
copies of the values 
\[
\Delta = H + \epsilon_1 {\mu \over 3} M^{12} + \epsilon_2 {\mu \over 6} M^{45} 
+ \epsilon_3 {\mu \over 6} M^{67} - \epsilon_2 \epsilon_3 {\mu \over 6} M^{89}
\]
and $\epsilon_i = \pm 1$ may be chosen independently. Highest weight BPS states 
will then be exactly those for which some of these eigenvalues vanish. It is now 
straightforward to describe the complete set of BPS states in the spectrum by 
inspecting table 2.

The eigenvalues of $12(H / \mu, M^{12}/3, M^{45}/6, M^{67}/6, M^{89}/6)$ for 
the individual oscillators (suppressing the block indices) are
\beas
\beta_{jm}: &&  (4j,4m, 0,0,0)\\ 
\alpha_{jm}: && (4j+4, 4m, 0,0,0)\\
x_{jm}: && (4j + 2 , 4m , \pm 2 ,0,0), \; 
 (4j+2, 4m, 0, \pm 2 , 0), \; 
(4j+2, 4m, 0 ,0, \pm 2) \\
\chi_{jm}: && ( 4j+3 , 4m, 1, 1, 1), \; (4j+3, 4m, 1, -1, -1) \\
&&(4j+3, 4m, -1 , 1, -1), \; (4j+3,4m,-1,-1,1) \\
\eta_{jm}: &&  ( 4j+1 , 4m, -1, -1, -1), \; (4j+1, 4m, 1, 1, -1) \\
&&(4j+1, 4m, 1 , -1, 1), \; (4j+1,4m,-1,1,1) 
\eeas
The $\{Q , Q\}$ eigenvalues are obtained by dotting these vectors
with
\[
{\mu \over 12} (1, \epsilon_1, \epsilon_2, \epsilon_3, - \epsilon_2 \epsilon_3) 
\; ,
\]
and a state corresponding to one of these oscillators is
invariant under the pair of supercharges corresponding to 
$(\epsilon_1, \epsilon_2, \epsilon_3)$ whenever the result is zero. It
is straightforward to check that this dot product is always greater
than or equal to zero (guaranteed by the BPS bound), and that it may only be 
zero 
for $m = \pm j$. For this choice, we find that each $\beta$, $\eta$, $x$, 
$\chi$, and $\alpha$ 
oscillator commutes with 4,3,2,1, and 0 pairs of supercharges,
respectively. 
\footnote{Note that the mode $x_{00}$ is an exceptional case and the number 
of pairs of supercharges preserved is 4.}

The oscillators which preserve each supercharge are
given by
\bea
+++: && \beta_{jj}, \;  x^{4-i5}_{jj}, \;  x^{6-i7}_{jj}, \; x^{8+i9}_{jj}, \; 
\chi_{jj}^{--+}, \;  \eta^{---}_{jj}, \;  \eta_{jj}^{-++}, \;
\eta_{jj}^{+-+}\cr \cr
++-: &&  \beta_{jj}, \;  x^{4-i5}_{jj}, \;  x^{6+i7}_{jj}, \; x^{8-i9}_{jj}, \; 
\chi_{jj}^{-+-}, \;  \eta^{---}_{jj}, \;  \eta_{jj}^{-++}, \;
\eta_{jj}^{++-}\cr \cr
+-+: && \beta_{jj}, \;  x^{4+i5}_{jj}, \;  x^{6-i7}_{jj}, \; x^{8-i9}_{jj}, \; 
\chi_{jj}^{+--}, \;  \eta^{---}_{jj}, \;  \eta_{jj}^{+-+}, \;
\eta_{jj}^{++-}\cr \cr
+--: &&  \beta_{jj}, \;  x^{4+i5}_{jj}, \;  x^{6+i7}_{jj}, \; x^{8+i9}_{jj}, \; 
\chi_{jj}^{+++}, \;  \eta^{-++}_{jj}, \;  \eta_{jj}^{+-+}, \;
\eta_{jj}^{++-} \label{list}
\eea
Here, we have labeled fermion oscillators by the signs of their $(M^{45},M^{67},
M^{89})$ eigenvalues and denoted $x^5 + i x^6$ by $x^{5+i6}$. The
values on the left denote the sign of $\epsilon_1$, $\epsilon_2$, and 
$\epsilon_3$ for a given pair of supercharges. The oscillators that
commute with supercharges corresponding to $\epsilon_1 = -1$ are given
by an identical table with the signs on the left reversed and $m= -j$ for all 
oscillators instead of $m=j$.

Since all of the quantum numbers are additive, a state obtained by
acting with an arbitrary combination of oscillators in a single row of
the list (\ref{list}) will be invariant under the corresponding
supercharge. Thus, up to $SO(3) \times SO(6)$ rotations, all BPS
states in the theory may be obtained by acting on the Fock space
vacuum with an arbitrary combination of oscillators in any given row
above. States for which all of the oscillators appear simultaneously
in $k$ rows of the table above will be left invariant by $2k$
supercharges. For example, a general class of 1/4 BPS states 
preserving 8 supercharges is
\[
\prod_j (b^\dagger_{jj})^{n_j}| 0 \rangle \;. 
\]
where $b$ is the creation operator associated with $\beta$.
It is interesting to note that the simple 1/4 BPS states, 
\[
(b^\dagger_{22})^n | 0 \rangle
\]
correspond classically to the rotating ellipsoidal membrane
state found by Bak \cite{bak}.

We have thus given the complete set of BPS states of the matrix model
for the $\mu= \infty$ limit. It would be interesting to determine
which of these remain BPS as we turn on interactions (at finite
$\mu$). This issue is discussed further in the remarks of section 8.

\section{Perturbation theory}

In this section, we move away from the $\mu = \infty$ limit and study 
perturbation theory about the free theory described in the previous section. 
Specifically, we would like to determine the range of parameters ($N,R,\mu$) and 
energies for which perturbation theory is valid. As a first step, we rewrite the 
interaction Hamiltonian in terms of normalized oscillators to determine the 
value of the ``bare coupling'' corresponding to different vacua. We then perform 
explicit calculations of the energy shift for the first non-trivial excited 
states about the $X=0$ vacuum and the irreducible vacuum to determine the actual 
effective coupling controlling the perturbation expansion. Since these
states are non-BPS, we expect their energy shifts to be generic. For
fixed values of the parameters, we then estimate the height of the
potential separating different vacua. This corresponds to the energy
above which perturbation theory becomes inapplicable.

\subsection{Expansion in terms of normalized oscillators}

In the previous section, we have explicitly diagonalized the quadratic actions, 
writing them in terms of canonically normalized oscillators. It is useful to 
consider expanding the interaction Hamiltonian in terms of these 
oscillators, in 
order to more accurately determine the parameter in the action suppressing 
interaction terms. 

In general, the cubic interactions in the matrix Lagrangian take the form
\[
H_3 \sim \left({R \over \mu} \right)^{3 \over 2} \tr(A_1 [A_2,A_3]) 
\]
while the quartic interactions take the form
\[
H_4 \sim \left({R \over \mu} \right)^3 \tr \left([A_1, A_2] [A_3, A_4] \right)
\]
where $A_i$ represent any of our matrix variables.

As an example, we consider a vacuum corresponding to $K$ copies of the $N_1$ 
dimensional irreducible representation, such that $K N_1 = N$. In this case, as 
discussed in section 5.3, it is convenient to split the matrices $A$ into $N_1 
\times N_1$ blocks, and we find from the analysis of section 5.2 that the 
individual blocks may be written in terms of normalized oscillators $A_{jm}$ as  
\be
\label{Aexp}
A^{kl} = {1 \over \sqrt{N_1}} c_{jm} A^{kl}_{jm} Y^{N_1}_{jm}
\ee
where $c_{jm}$ are coefficients independent of $N$. To evaluate the coefficients 
in the cubic and quartic actions, we note (see, for example \cite{hoppe,iktw})
\be
\label{Ycomm}
[Y^N_{j_1 m_1}, Y^N_{j_2 m_2}] = {1 \over N} b_{j_1 m_1 \; j_2 m_2 \; j_3 m_3} 
(Y^N_{j_3 m_3})^\dagger
\ee
where $b$ is zero when $j_1+j_2+j_3$ is even and
otherwise
\beas
b_{j_1 m_1 \; j_2 m_2 \; j_3 m_3} &=& (-1)^N 2N^{3 \over 2}  \sqrt{(2j_1+1)(2j_2 
+ 1)(2j_3 + 1)}\\
&& \qquad \qquad \left( \ba{ccc} j_1 & j_2 & j_3 \\ m_1 & m_2 & m_3 \ea \right) 
\left\{ 
\ba{ccc} j_1 & j_2 & j_3 \\ {N-1 \over 2} & {N-1 \over 2} & {N-1 \over 2} \ea 
\right\}\ .
\eeas
Explicit formulae for the Wigner 3j and 6j symbols appearing here may be found 
in \cite{edmonds}. Using these, it is straightforward to show that $b$ has 
a finite 
limit 
as $N \to \infty$ with corrections of order $1/N$. 

Using (\ref{Aexp}) and (\ref{Ycomm}) and recalling the normalization condition 
(\ref{Ynorm}) for $Y$'s, we may now rewrite the interaction Hamiltonian in 
terms 
of normalized oscillators. We find schematically
\beas
H_3 \sim \left({R \over \mu N_1} \right)^{3 \over 2} b c c c AAA \\
H_4 \sim \left({R \over \mu N_1} \right)^{3 \over 2} b b c c c c A A A A 
\eeas
For large $N$ the leading terms in the $b$ and $c$ coefficients are independent 
of $N$, so we conclude that the coupling corresponding to the vacuum with $K$ 
spheres of radius $r = (\mu N_1 / 6R)$ is given by
\[
g = \left({R \over \mu N_1} \right)^{3 \over 2}  \propto {1 \over r^{3 \over 2}}
\]
For a fixed value of $N$, we see that the irreducible vacuum appears to have the 
weakest coupling, $g_{irr} = (R /(\mu N))^{3 \over 2}$ while the $X=0$ vacuum is 
the most strongly coupled, with $g_{X=0} = (R/ \mu)^{3 \over 2}$.

In systems with many degrees of freedom, it is often the case that the 
perturbation expansion is controlled by some effective coupling which is 
different than the bare coupling appearing in the Lagrangian. To determine 
the 
true effective coupling in our case, we now turn to some explicit perturbative 
calculations.

\subsection{$X=0$ vacuum}

The $X=0$ vacuum is invariant under the full $U(N)$ symmetry group of the model. 
As discussed in section 5.4, it is then convenient to introduce matrix creation 
and annihilation operators transforming in the adjoint of this symmetry group,
\beas
X^i &=& \sqrt{3 \over 2}(A_i + A_i^\dagger)\\
\Pi^i &=& -{ i \over \sqrt{6}} (A_i - A_i^\dagger)\\
X^a &=& \sqrt{3} (A_a + A_a^\dagger)\\
\Pi^a &=& - {i \over \sqrt{12}}(A_a - A_a^\dagger)
\eeas
in terms of which the quadratic Hamiltonian becomes
\[
H_2 = \tr({1 \over 3} A_i^\dagger A_i + {1 \over 6} A_a^\dagger A_a + {1 \over 
4} \psi^{\dagger I \alpha} \psi_{I \alpha} ) \; .
\]
Physical states satisfying the Gauss law constraint will then be obtained by 
acting with traces of products of these matrix creation operators on the Fock 
space vacuum, for example
\[
\tr(A_a^\dagger A_b^\dagger \psi^{\dagger I \alpha}) \tr(A_i^\dagger 
A_c^\dagger) |0 \rangle \; .
\]
The lowest lying states are given in table 3.
\begin{table}[ht]
\begin{center}
\begin{tabular}{|c|c|c|c|}
\hline
{\rm State} &  $SO(6) \times SO(3)\ $   {\rm Rep.} & {\rm Energy} & {\rm 
Degeneracy} \\ \hline
$|0 \rangle$ & $(1,1)$ & $0$ &$ 1$ \\ 
$\tr(A_a^\dagger) |0 \rangle$ & $(6,1)$ & ${\mu \over 6}$ & $6$ \\
$\tr(\psi^{\dagger I \alpha}) |0 \rangle$ & $(\bar{4},2)$ & ${\mu \over 4}$ & 
$8$ \\
$\tr(A_i^\dagger) |0 \rangle$ & $(1,3)$ & ${\mu \over 3}$ & $3$ \\
$\tr(A_a^\dagger) \tr(A_b^\dagger) |0 \rangle$ & $(1,1)+(20,1)$ & ${ \mu 
\over 3}$ & $1 + 20$ \\ 
$\tr(A_a^\dagger A_b^\dagger)- {1 \over N} \tr(A_a^\dagger) 
\tr(A_b^\dagger) |0 \rangle$ & $(1,1) + (20,1)$ & ${ \mu \over 3}$ & 
$1 + 20$ \\
\hline
\end{tabular} \caption{Lowest energy states about the $X=0$ vacuum}
\end{center}
\end{table}

We can now investigate the change in energy of these states upon turning on the 
interaction Hamiltonian, using perturbation theory in ${R \over \mu}$. Note 
that 
only the last state in the list above can get a shift in energy, since the other 
states are excitations in the $U(1)$ part of the theory, which is free. 

It is easy to see that the energy shift at first order in perturbation theory
\[
\langle \psi | H_3 | \psi \rangle
\]
will be zero for all states, since all terms in $H_3$ have non-zero $H_2$ 
eigenvalue. Thus $H_3$ acting on any state gives a combination of basis states
all of whose energies are different than that of the original state.

At second order in perturbation theory, the energy shift is given by the usual 
formula,
\bea
\Delta &=& \Delta_4 + \Delta_3 \nonumber\\
&=& \langle \psi | H_4 | \psi \rangle + \sum_n {1 \over E_\psi - E_n} |\langle n 
| H_3 | \psi \rangle |^2 \label{2pert3}
\eea

Before calculating the energy shift for the excited state, it is useful to 
verify explicitly that the vacuum state receives no correction. In appendix C.1, 
we show that for the vacuum state, $\Delta_4 = -\Delta_3 = {891 \over 4}(N^3 - 
N)$, so the vacuum state does remain at zero energy to second order in 
perturbation theory. The vacuum wavefunction does receive a correction at first 
order in perturbation theory, given in equation (\ref{vacshift}).

We now compute the vacuum shift for the lowest energy non-trivial state (i.e. 
involving more than $U(1)$ oscillators),
\[
|\psi \rangle = {1 \over \sqrt{12(N^2-1)}} (\tr(A_a^\dagger A_a^\dagger)- {1 
\over N} \tr(A_a^\dagger) \tr(A_a^\dagger) ) |0 \rangle 
\]
This normalized state has been chosen to be orthogonal to the state
\[
\tr(A_a^\dagger) \tr(A_a^\dagger) ) |0 \rangle
\]
which cannot receive any energy correction.

The calculation of the energy shift to second order in perturbation theory is 
performed in appendix C.1, and we find 
\[
\Delta = 108 N \mu \left( {R \over \mu} \right)^3  
\]
Thus, the effective coupling controlling the perturbation expansion appears to 
be (ignoring numerical factors)
\[
N (R/ \mu)^3  = g^2 N
\] 
where $g = (R / \mu)^{3 \over 2}$ was the bare coupling suppressing the 
interaction terms in the action near $X=0$.

\subsection{Perturbation theory about irreducible vacuum}

We next consider perturbation theory for the single membrane vacuum, 
corresponding to the irreducible representation of $SU(2)$. Here, the low-energy 
states are obtained by acting with creation operators corresponding to the 
oscillators in table 1. The states with lowest energy are summarized in table 4, 
where we have defined $a_{jm}^a$, $a_{jm}$, and $b_{jm}$ to be canonical 
annihilation operators associated with the oscillators $x_{jm}^a$, 
$\alpha_{jm}$, and $\beta_{jm}$ respectively. 

\begin{table}[ht]
\begin{center}
\begin{tabular}{|c|c|c|c|}
\hline
{\rm State} &  $SO(6) \times SO(3)\ $   {\rm Rep.} & {\rm Energy} & {\rm 
Degeneracy} \\ \hline
$|0 \rangle$ & $(1,1)$ & $0$ &$ 1$ \\ 
$ (a^a_{00})^\dagger|0 \rangle$ & $(6,1)$ & ${\mu \over 6}$ & $6$ \\
$ \eta^{I \dagger}_{{1 \over 2} \; m} |0 \rangle$ & $(\bar{4},2)$ & ${\mu \over 
4}$ 
& $8$ \\
$ b^\dagger_{1m} |0 \rangle$ & $(1,3)$ & ${\mu \over 3}$ & $3$ \\
$ (a_{00}^a)^\dagger (a_{00}^b)^\dagger |0 \rangle$ & $(1,1)+(20,1)$ & ${ \mu 
\over 3}$ & $1 + 20$ \\ 
$ a^\dagger_{00} |0 \rangle$ & $(1,1)$ & ${ \mu \over 3}$ & 
$1$ \\
\hline
\end{tabular} \caption{Lowest energy states about the irreducible vacuum}
\end{center}
\end{table}

Note that $x^a_{00}$, $\beta_{1m}$ and $\eta_{{1 \over 2} \; m}$ correspond to 
oscillators in the free $U(1)$ part of the theory, so it is only the last state 
that can receive an energy shift in perturbation theory. This is the breathing 
mode of the sphere. 

In appendix C, we compute the leading energy shift for an arbitrary occupation 
number of this breathing mode, i.e., for the state
\[
{1 \over \sqrt{n!}} (a^\dagger_{00})^n |0 \rangle \; . 
\]
The result, coming at second order in perturbation theory from the terms 
(\ref{2pert3}) is
\beas
\Delta / \mu &=& - 216 \left( \ba{c} n \\ 2 \ea \right) {R^3 \over \mu^3 N(N^2-
1)}  \\
&=& -216  \left( \ba{c} n \\ 2 \ea \right)  g^2  (1 + {\cal O} ({1 \over N}) 
)\ . 
\eeas
(Note that for $n=1$ the energy shift $\Delta$ is zero.)
In this case, we see that the leading correction for low energy states is 
controlled by the ``bare'' coupling $g^2$ without any additional $N$-dependence 
and is therefore finite in the large $N$ limit.

\subsection{Validity of perturbation theory}

In the previous sections, we have computed the leading correction to the 
energies of the first nontrivial excited states above the irreducible 
single-membrane vacuum and the $X=0$ vacuum. We found that the effective 
coupling controlling the perturbative corrections is given by $\lambda = g^2 K$, 
where $K=1$ and $g^2 = ({R \over \mu N})^{3 \over 2}$ for the single
membrane case and $K=N$ and $g^2 = ({R \over \mu})^{3 \over 2}$ for
the $X=0$ vacuum.\footnote{The other vacua are characterized by a 
couplings intermediate between these values.} Note that $K$
corresponds to the number of coincident membranes characterizing the
vacuum and in this sense $\lambda$ behaves like a 't Hooft coupling. 

In situations where perturbative corrections are small for low lying states, 
we would now like to estimate the energy above which perturbation theory breaks 
down. In terms of the classical potential, we expect this to occur when we move 
far enough away from the minimum so that the interaction terms in the potential 
become of the same order of magnitude as the quadratic terms. 

To obtain an initial, crude estimate of this energy scale, consider starting at 
this irreducible vacuum and exciting the breathing mode considered in the 
previous section, that is, take $Y^i = \alpha {\mu \over 3R} J^i$. Plugging this 
in to the classical potential, we find
\beas
V &=& {1 \over 2} R \tr(J^i J^i) \left( {\mu \over 3R} \right)^4 \alpha^2 (1 + 
\alpha)^2 \\
&=& {R N (N^2 - 1) \over 8} \left( {\mu \over 3R} \right)^4 (\alpha^2 + 2 
\alpha^3 + \alpha^4)
\eeas
Thus, we see that the cubic and quartic terms in the potential will be small 
compared to the quadratic term when $\alpha \ll 1$. In terms of the energy, this 
translates to the condition (ignoring numerical factors)
\[
{E \over \mu } \ll \left({\mu N \over R}\right)^3 = {1 \over g^2} \; .
\]
Our crude estimate suggests that the quadratic approximation to the potential is 
good for energies less than $\mu / g^2$. 

We expect this estimate to be true for a generic direction away from the minimum 
However, it turns out that for certain paths away from the minimum, the 
interaction terms can become important at a much lower energy. In appendix D, we 
provide an explicit path between the irreducible vacuum labeled by $N$ and the 
nearby vacuum labeled by $(N-1,1)$ along which the maximum energy is of order 
$(\mu N / R)^3 {1 \over N} = {1 \over g^2 N}$ rather than $1/g^2$.\footnote{In 
terms of the natural metric $\tr((X^A)^2)$ on the space of matrices, this choice 
for the final state represents the closest vacuum.} This path corresponds to 
the discretized version of a continuum picture in which a thin tube of membrane 
extends from the south pole of the sphere to the origin where a small second 
sphere can grow from the tip of the tube, as depicted in figure 2.\footnote{We 
thank Washington Taylor for a valuable discussion about this.} 

\begin{figure}
\centerline{\epsfysize=0.95truein \epsfbox{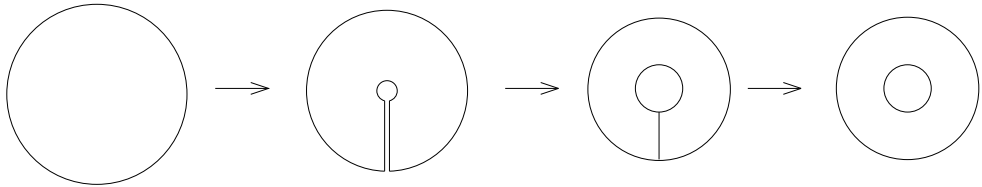}}
%\vskip -0.4 cm
 \caption{Low energy path between nearby vacua}
\label{sph}
\end{figure}

Thus, our crude estimate for the energy at which interaction 
terms become important was in fact too large by a factor of $N$. At least along 
a specific direction away from the minimum, we find that interactions must be 
important at an energy 
\[
{E \over \mu} \sim {1 \over N} \left({\mu N \over R} \right)^3 = {1
 \over g^2 N} 
 \; .
\]
From this result, one might conclude that the condition for a
perturbative regime to exist should be obtained by demanding that this
energy is significantly greater than the energy of the lowest lying
modes, that is, $g^2 N \ll 1$. On the other hand, we have seen
explicitly that the leading perturbative correction for certain low
energy states was small as long as $g^2 \ll 1$, a much weaker
condition. It is possible that the low energy paths between vacua we
have found are very narrow for large $N$ (a set of very small measure
in the space of paths away from the minimum) and therefore that
perturbation theory may be valid for larger energies than the height
of these paths would indicate. In particular, from the calculation of 
appendix C.2, we see that the the leading perturbative correction to
the breathing mode with occupation number $n$ is small when $n^2
g^2 \ll 1$, in other words when $E \ll \mu /g$. Further work will be required to 
decisively determine which of these conditions (or some intermediate condition) 
actually governs the validity of perturbation theory.

\section{M-theory limit}

In this section, we discuss the implications of our results for the M-theory 
limit of the matrix model.

\subsection{Implications of perturbative results}

M-theory on the pp-wave background should be described using the matrix model by 
taking the $N \to \infty$ limit with fixed $p^+ = N/R$ and fixed $\mu$. The 
resulting theory is characterized by the single boost-invariant dimensionless 
parameter $\mu p^+ l_p^2$. In this background, a single membrane with momentum 
$p_i^+$ in the pp-wave background will expand into a spherical BPS ``giant 
graviton'' configuration with radius $\mu p_i^+$ \cite{bmn}. In general, the 
total momentum $p^+$ may be divided among any number of membranes, and each of 
these membranes will have a radius proportional to its momentum. This is 
precisely the picture we get from the large $N$ limit of the matrix model.

In section 5, we calculated the spectrum of the matrix model in the limit of 
large $\mu$. We found that the individual oscillator masses were independent of 
$N$. For the irreducible vacuum, $N$ acts as a cutoff controlling the maximum 
allowed frequency of the modes. Thus, in the large $N$ limit, we find a well-
defined spectrum, which we may associate with the spectrum of the free part of 
the field theory on a single spherical supermembrane in the pp-wave background. 

For reducible representations involving $k$ copies of a single irreducible 
representation, we obtain $k^2$ copies of the spectrum for the corresponding 
irreducible representation, suggesting a non-abelian structure for coincident 
membranes. For reducible vacua involving different irreducible representations 
of dimension $N_1$ and $N_2$, corresponding to non-coincident membranes, we 
found that the spectrum corresponding to off-diagonal degrees of freedom 
connecting the two representations has a lower bound on oscillator masses 
proportional to ${\mu \over 2}|N_1 - N_2|$. Thus, if we fix the radius 
parameters $(\mu N_i/R)$ of the two membranes and take the large $N$ limit, we 
find that all of the off-diagonal degrees of freedom develop infinite masses, 
going like $N(p_1^+ - p_2^+)/p^+$.   

It is not clear whether the M-theory limit is compatible with a valid 
perturbation expansion. In explicit perturbative calculations, we found that the 
leading corrections to the energies of low-lying states above the 
single-membrane vacuum are finite in the the M-theory limit and small when $\mu 
p^+ \gg 1$. On the other hand, we have shown in appendix D that there exist 
paths in configuration space between any two vacua with maximum energy of order 
$\mu (\mu p^+)^3 / N$, which goes to zero in the M-theory limit. This indicates 
that in the large $N$ limit, the classical vacua are no longer isolated, but 
rather are all connected by finite, zero-energy flat directions in the 
potential. Physically, these flat directions correspond to decreasing the radius 
(and light-cone momentum) of any membrane while increasing the radius of any 
other membrane at the same rate (or growing a new membrane at the origin). 
Quantum mechanically, these flat directions would seem to spoil any perturbative 
study of the model, since none of the classical vacua are long-lived and the 
low-energy wave-functions would presumably spread out over all the classical 
vacua. 
   
\subsection{The supermembrane field theory}

From the supermembrane point of view, the matrix model can be thought of as a 
regulated version of the light-cone supermembrane field theory on some fixed 
topology worldvolume. When interpreted in this context, our results have a 
number of interesting implications. 

To be definite, consider starting from the supermembrane Hamiltonian 
(\ref{membranetotal}), and choose a spherical worldvolume of radius 1 labeled 
by coordinates $x_1,x_2,x_3$ with $x_1^2 + x_2^2 + x_3^2 =
1$.\footnote{In terms of the usual spherical coordinates $\phi,
\theta$, we have $x_3 = \cos\theta$, $x_1 = \sin\theta \cos\phi$,
and $x_2 = \sin\theta \sin\phi$ and the Poisson bracket on the
sphere is defined by $\{\theta, \phi \} = {1\over \sin \theta}$.}  
The potential term in the bosonic part of the Hamiltonian is 
\be
V \propto \left(\epsilon^{ijk}\{X^i, X^j\}-{2\mu p^+\over 3} X^k\right)^2+
\{X^a, X^b\}^2 +2\{X^i, X^a\}^2+\Big({\mu p^+\over 3}\Big)^2 {(X^a)}^2\ .
\ee
where $\{A, B\} = \epsilon^{ijk} x_i \partial_j A \partial_k B$ is the canonical 
Poisson bracket. This potential has a minimum when $X^a=0$ and  
\be\label{Sphere}
\{X^i, X^j\}=\epsilon^{ijk}{\mu p^+\over 3} X^k
\ee  
In the continuum case, there are only two vacua corresponding to non-singular 
field configurations, namely $X=0$ and $X^i = R_{sph} x^i$ where $R_{sph} = {\mu 
p^+ / 6}$. 

By analogy with the matrix model analysis above, one might have been interested 
to study the field theory expanded about the large membrane solution. The 
resulting theory has an unusual gauge symmetry coming from the symmetry under 
area-preserving diffeomorphisms,
\[
\delta Y^i = -R_{sph} \epsilon^{ijk} x^j \partial^k \Lambda + \{Y^i , 
\Lambda \} 
\]
One may attempt to decompose $Y$ into fields with more standard gauge 
transformation properties as
\be
\phi={1\over R_{sph}}\ x \cdot \vec{Y}\ ,\ \ \
\vec A={1\over R_{sph}}\ x \times \vec{Y}\ ,
\ee
To leading order, the fields $A$ and $\phi$ have the usual transformation 
properties of a gauge field and a scalar, and furthermore $A$ appears in the 
action as a massless gauge field while $\phi$ is massive. However, there are 
also nonlinear terms in the gauge transformation which may only be removed by 
adding higher order terms in the definition of $A$ and $\phi$. These introduce 
even higher order terms in the gauge transformation, and so forth, indicating 
that the definition of a proper gauge field with transformation property $\delta 
A_i = \partial_i \Lambda - \{A_i, \Lambda\}$ would require an infinite 
series of 
terms in $Y$. Thus, for now we shall consider the theory written in terms of 
$Y$.

At first sight, this theory looks non-renormalizable, since the $\{ \; ,\; \}^2$ 
terms have dimension 6, suggesting for example that one would find divergent 
one-loop contributions in the quadratic effective action going like $\Lambda^5$. 
In the matrix regularized theory, the role of the UV cutoff is played by $N$, so 
we might have expected $N^5$ behavior in the mass shift we have calculated for 
the breathing mode. In reality, our result is ${\cal O}(1)$ in the large $N$ 
limit, indicating that the theory may have much better renormalizability 
properties than it appears to. This is at least partly due to supersymmetric 
cancellations. Another possible explanation is that we should really consider 
the theory as a $\theta \to 0$ limit of a noncommutative 2+1 dimensional field 
theory on a sphere with only quartic interactions. Such a theory would be 
expected to give only linear divergences in the quadratic effective action, or 
even weaker behaviour in the supersymmetric case. It would be quite interesting 
if the Poisson bracket field theory turned out to be perturbatively 
renormalizable.

We noted that the supermembrane field theory has two nonsingular classical 
vacuum solutions, however, from the matrix theory point of view, we have seen 
that the theory has a continuum of other vacua connected to these by finite 
zero-energy flat directions. In the continuum theory, these vacua will appear as 
singular field configurations. To understand what these look like, recall that 
one path between the single membrane vacuum and a vacuum with two spheres 
involved forming a thin tube of membrane from the south pole of the original 
sphere which moves up to the origin and nucleates another sphere. From the 
membrane field theory point of view, this corresponds to nucleating a small 
bubble of domain wall at some point on the sphere, where the inside of the 
domain wall maps to the new sphere, and the domain wall itself maps to the tube. 
In the final vacuum state with two spheres, the tube has shrunk to zero size, so 
the domain wall becomes singular. Thus, the flat directions we have found 
correspond to the existence of massless singular domain walls, or tensionless 
strings, in the supermembrane field theory. Presumably, these nonperturbative 
massless excitations make the theory extremely difficult to study.

\section{Summary and Discussion}

In this paper, we have studied the matrix model proposed by Berenstein, 
Maldacena, and Nastase to describe M-theory on the maximally supersymmetric pp 
wave solution of eleven-dimensional supergravity. We have seen that the model 
may be derived directly as a regularized form of the light-cone supermembrane 
action in the pp-wave background, or alternatively (with certain assumptions) 
from D0-branes in a corresponding type IIA background. 

We noted that for large values of the mass parameter $\mu$, the theory may be 
expanded about its classical vacua to give a quadratic theory with interactions 
suppressed by powers of $1/\mu$. In the large $\mu$ limit for fixed $N$ and $R$, 
we determined the complete physical spectrum of the theory and found that the 
individual energy levels are rational multiples of $\mu$ independent of $N$. 
We characterized the complete set of BPS states in this limit, finding
infinite towers of states preserving 2,4,6, and 8 of the 32
supercharges (in addition to the 1/2 BPS vacuum states).
We then considered perturbation theory about the vacua, and found that vacua 
corresponding to more spheres at smaller radii are generally more strongly 
coupled. For the most weakly coupled vacuum, corresponding to a single membrane 
sphere, we found that the leading corrections to the energies of low-lying 
excited states were small when $g^2 = 1/(\mu p^+)^3 \ll 1$, independent of $N$. 
On the other hand, we saw that 
the various vacua are connected by paths along which the maximum energy is $(\mu 
p^+)^3/N$ indicating a more restricted regime of validity for perturbation 
theory incompatible with the M-theory limit.

A very interesting direction to pursue would be to establish which of
the BPS states appearing in the $\mu= \infty$ limit remain
supersymmetric for finite $\mu$. Typically, BPS multiplets may only
become non-BPS by combining with other BPS multiplets to form larger
non-supersymmetric representations. By better understanding the
representation theory of the pp-wave superalgebra, it may be possible
to show that certain BPS multiplets are protected. The existence of
protected states would provide reliable information about the spectrum
even in the regime where perturbation theory is invalid. Another
approach to determining which states remain BPS would be to perform a
direct calculation of the energy shift
in perturbation theory, as we have done in this paper for two simple non-BPS
states. Since the energy of BPS states is determined exactly by their
angular momenta, any state which remains BPS when interactions are
turned on must have zero energy shift in perturbation theory. We note
that at least one of the 1/4 BPS states we have found at $\mu=
\infty$ should remain BPS, namely the rotating ellipsoidal membrane
solution considered in \cite{bak}. 

We would now like to briefly address the issue of vacuum 
stability. We have seen that classically, and to leading orders in perturbation 
theory, the various vacua are all degenerate and have zero energy. It is 
not clear whether this will be true nonperturbatively in the full 
quantum theory. We know that for $\mu = 0$ there exists precisely one 
normalizable zero-energy state, however it is possible that other zero-energy 
states (corresponding to multiple gravitons) exist which become non-normalizable 
as $\mu \to 0$. If the classical vacua do not remain degenerate, then the exact 
wavefunctions of the theory will presumably not be localized around any one of 
the classical vacua, and transitions will occur. On the other hand, if $\mu$ is 
large, the low-energy wavefunctions we have been discussing should serve as 
useful approximate wavefunctions over time scales much shorter than the typical 
decay rate from one vacuum to another. We expect that for large values
of $\mu$, the path integral governing the transition rate between different 
minima 
of the classical potential (corresponding to membrane transition amplitudes) 
will 
be dominated by classical instanton solutions so that the rates may be estimated 
by a stationary phase approximation. Some examples of these instanton solutions 
have been worked out in \cite{bhp}.

It is interesting to note that the more commonly studied case of type IIB string 
theory on the maximally supersymmetric pp-wave solution of type IIB 
supergravity should behave very much like the M-theory case when the
string coupling is taken to be of order 1. In particular, along the
range of $\mu p^+ \alpha'$  with fixed $g_s = 1$, the 
classical vacua of the theory are characterized by collections of D3-brane giant 
gravitons with the sum of the squared radii proportional to $\mu p^+$ (in units 
where $\alpha' = 1$). It would be interesting to determine whether a matrix 
model for the DLCQ of type IIB string theory on this background (discussed 
recently in \cite{gopakumar}) also contains a 
perturbative regime for large $\mu$. 

\vskip 1cm

\noindent
{\large \bf Acknowledgments}\\
\\
We would like to thank Renata Kallosh, Shiraz Minwalla, Stephen Shenker, and 
Washington Taylor for useful discussions. The research of K.D. is supported in 
part by a David and Lucile Packard Foundation Fellowship 2000-13856. The work of 
M. M. Sh-J., and M.V.R is supported in part by NSF grant PHY-9870115 and in part 
by funds from the Stanford Institute for Theoretical Physics.

\vskip 1cm

\appendix

\section{Fermions in $SU(2) \times SU(4)$ notation.}

\label{fermions}

The massive matrix model preserves an $SO(3) \times SO(6) \sim SU(2) \times 
SU(4)$ subgroup of the original $SO(9)$ symmetry of the BFSS model.
It is convenient to write the action in a form for which 
the $SO(3) \times SO(6)$ transformation properties of all fields are manifest. 
This is already the case for the bosons, so we turn to the fermionic fields. 
Under the decomposition
\[
SO(9) \to SO(6) \times SO(3)
\] 
the spinor representation splits up as
\beas
{\bf 16} &\to& {\bf (4,2)} + {\bf (\bar{4}, \bar{2})} \\
\Psi &\to& \psi_{I \alpha}, \tilde{\psi}^{J \beta}
\eeas
where we take $I$ as a fundamental $SU(4)$ index and $\alpha$ as a fundamental 
$SU(2)$ index. Our spinors obey a reality condition, which in the reduced 
notation becomes simply
\[
(\psi^\dagger)^{I \alpha} = \tilde{\psi}^{I \alpha}
\]
To decompose the terms in the action, we introduce the matrices ${\sf
g}^a_{IJ}$ which relate the antisymmetric product of two ${\bf 4}$ 
representations of $SU(4)$ to the vector of $SO(6)$. Then, we may 
choose\footnote{In this representation, the Dirac matrices and fermions are 
not real, however the spinors still obey a reality condition which reduces 
them to 16 independent real components. }
\[
\Psi = \left( \ba{c} \psi_{I \alpha}  \\ \epsilon_{\alpha \beta} 
\psi^{\dagger I \beta} \ea \right)
\]
and
\[
\gamma^a = \left( \ba{cc} 0 & 1 \times {\sf g}^a \\ 1 \times ({\sf 
g}^a)^\dagger & 0 \ea \right) \qquad \gamma^i = \left( \ba{cc} -\sigma^i \times 
1 & 0  \\ 0 & \sigma^i \times 1 \ea \right)
\]
These Dirac matrices satisfy the Clifford algebra as long as we take 
normalizations so 
that
\[
{\sf g}^a ({\sf g}^b)^\dagger + {\sf g}^b ({\sf g}^a)^\dagger = 2 
\delta^{ab} 
\]
We may then rewrite the quadratic fermion terms in our original action 
as\footnote{Since we are working with fermions that are not explicitly real, we 
need to replace $\Psi^\top$ with $\Psi^\dagger$.}
\beas
{i \over 2} \Psi^\dagger D_0 \Psi  &\rightarrow& i \psi^{\dagger I
\alpha} D_0 \psi_{I \alpha} \\
- { i \over 8} \Psi^\dagger \gamma^{123} 
\Psi  &\rightarrow& - {1 \over 4} 
\psi^{\dagger I \alpha} \psi_{I \alpha} \\
{1 \over 2} \Psi^\dagger \gamma^i [X^i, \Psi] &\rightarrow& -\psi^{\dagger I 
\alpha} \sigma^i_\alpha {}^\beta [X^i, \psi_{I \beta}]\\
{1 \over 2} \Psi^\dagger \gamma^a [X^a, \Psi] &\rightarrow& {1 \over 2} 
\epsilon_{\alpha \beta} \psi^{\dagger \alpha I} {\sf g}^a_{IJ} [X^a, 
\psi^{\dagger \beta J}] - {1 \over 2} \epsilon^{\alpha \beta} \psi_{\alpha I} 
({\sf g}^\dagger)^{a IJ} [X^a, \psi_{\alpha J}]\ .
\eeas

\section{Symmetry algebra in terms of matrix generators.}

In this appendix, we review the symmetry algebra of the matrix model and provide 
explicit expressions for the generators in terms of the matrix model variables. 

The bosonic generators include the 
Hamiltonian $H$, the light-cone momentum $P^+$ (a central term in the algebra) 
and rotation generators of the $SO(3) \times SO(6)$ symmetry which we denote by 
$M^{ij}$ and $M^{ab}$ respectively. In addition, there are generators $a_a$ and 
$a_i$ with nonvanishing commutation relations
\beas
\left[ a_a , a_b^\dagger \right] &=& P^+ \delta_{ab} \\ 
\left[ a_i , a_j^\dagger \right] &=& P^+ \delta_{ij} \\
\left[ H , a_i \right] &=& - {\mu \over 3} a_i\\
\left[ H , a_a \right] &=& - {\mu \over 6} a_a\\
\left[ M^{ab} , a_c \right] &=& i \delta_{ac} a_b - i \delta_{bc} a_a\\
\left[ M^{ij} , a_k \right] &=& i \delta_{ik} a_j - i \delta_{jk} a_i\\
\eeas
These are the creation and annihilation operators corresponding to the decoupled 
$U(1)$ part of the theory which describes a particle (the center of mass 
coordinate) in a harmonic potential.

The algebra has in addition 32 fermionic generators. The first 16, labeled by 
$q_{I \alpha}$ change the overall polarization state in the $U(1)$ part of the 
theory, while the other $16$, labeled by $Q_{I \alpha}$ act non-trivially.

These generators obey anticommutation relations
\beas
\{ Q^{\dagger I \alpha}, Q_{J \beta} \} &=& 2 \delta^I_J \delta^\alpha_\beta H   
- {\mu \over 3} \epsilon^{ijk} \sigma^k_{\beta} {}^\alpha \delta^I_J
M^{ij} - {i \mu \over 6} \delta^\alpha_\beta ({\sf g}^{ab})_J {}^I M^{ab}\\
\{q_{I \alpha}, Q_{J \beta} \} &=& - i \sqrt{\mu \over 3} {\sf g}^a_{IJ} 
\epsilon_{\alpha \beta} a_a \\
\{q^{\dagger I \alpha}, Q_{J \beta} \} &=& -i \sqrt{2\mu \over 3} 
\sigma^i_\beta {}^\alpha \delta^I_J a_i^\dagger\\
\{ q^{\dagger I \alpha}, q_{J \beta} \} &=& \delta^\alpha_\beta \delta^I_J P^+
\eeas
and have nonvanishing commutation relations with the bosonic generators given by
\beas
\left[H, q_{I \alpha} \right] &=& -{\mu \over 4} q_{I \alpha}\\ 
\left[H, Q_{I \alpha} \right] &=& {\mu \over 12} Q_{I \alpha}\\
\left[Q_{I \alpha}, a_a^\dagger \right] &=& -i \sqrt{\mu \over 3} {\sf
g}^a_{IJ} \epsilon_{\alpha \beta} q^{\dagger J \beta}\\
\left[Q_{I \alpha}, a_i \right] &=& \sqrt {2 \mu \over 3} i
\sigma^i_\alpha {}^\beta q_{I \beta}
\eeas 
with additional commutation relations between the fermionic generators and $M$'s 
appropriate to the $SO(3) \times SO(6)$ transformation properties of the $q$s 
and $Q$s. 

These generators are realized explicitly in the matrix model as
\beas
P^+ &=& {1 \over R}\tr(1) \\
a_i &=& {1 \over \sqrt{R}} \tr( \sqrt{\mu \over 6R} X^i + \sqrt{3R \over 2 \mu} 
i \Pi^i)\\
a_a &=& {1 \over \sqrt{R}} \tr( \sqrt{\mu \over 12R} X^a + \sqrt{3R \over \mu} i 
\Pi^a)\\
M^{ij} &=& \tr(X^i \Pi^j - X^j \Pi^i + i \epsilon^{ijk} \psi^\dagger \sigma^k 
\psi)\\
M^{ab} &=& \tr(X^a \Pi^b - X^b \Pi^a + {1 \over 2} \psi^\dagger {\sf g}^{ab} 
\psi)\\
q_{I \alpha} &=& {1 \over \sqrt{R}}\tr(\psi_{I \alpha})\\
Q_{I \alpha} &=& \sqrt{R} \tr \left( (\Pi^a - i {\mu \over 6R} X^a) 
{\sf g}^a_{IJ} 
\epsilon_{\alpha \beta} \psi^{\dagger J \beta} - (\Pi^i + i {\mu \over 3R} X^i)
 \sigma^i_\alpha {}^\beta \psi_{I \beta} \right.\\
& & \left. + {1 \over 2} [X^i, X^j] \epsilon^{ijk} \sigma^k_\alpha {}^\beta 
\psi_{I \beta} - {i \over 2} [X^a, X^b] ({\sf g}^{ab})_I^J \psi_{J \alpha} +i 
[X^i, X^a] \sigma^i {\sf g}^a_{IJ} \epsilon_{\alpha \beta} \psi^{\dagger J 
\beta} \right)
\eeas
where ${\sf g}^{ab} = {1 \over 2} ({\sf g}^a {\sf g}^{\dagger b} - {\sf g}^b 
{\sf g}^{\dagger a})$. The Hamiltonian $H$ may be read off from the
action in section 4.1. Using the quantum commutation relations 
\beas
[X^i_{kl}, \Pi^j_{mn}] = i \delta^{ij} \delta_{kn} \delta_{lm} \qquad
[X^a_{kl}, \Pi^b_{mn}] = i \delta^{ab} \delta_{kn} \delta_{lm} \qquad \{
\psi_{kl}^{\dagger I \alpha} , (\psi_{J \beta})_{mn} \} = \delta^I_J 
\delta^\alpha_\beta
 \delta_{kn} \delta_{lm} 
\eeas
we may verify explicitly that the algebra given above is satisfied.

For convenience, we also provide the supersymmetry algebra in the original 
$SO(9)$ 
notation of BMN,
\beas
\{ Q_\alpha, Q_\beta \} &=& 2 \delta_{\alpha \beta} H - { \mu \over 3} 
(\gamma^{123} \gamma^{ij})_{\alpha \beta} M^{ij} + { \mu \over 6} (\gamma^{123} 
\gamma^{ab})_{\alpha \beta} M^{ab}\\
\{ Q_\alpha, q_\beta \} &=& i \sqrt{\mu \over 3} \left( \{ {1 \over 2} (1 - i 
\gamma^{123})\gamma^a \}_{\alpha \beta} a_a^\dagger -  \{ {1 \over 2} (1 + i 
\gamma^{123})\gamma^a \}_{\alpha \beta} a_a \right)\\
&& -i \sqrt{2 \mu \over 3} \left( \{ {1 \over 2} (1 - i \gamma^{123})\gamma^i 
\}_{\alpha \beta} a_i -  \{ {1 \over 2} (1 + i \gamma^{123})\gamma^i 
\}_{\alpha \beta} a^\dagger_i \right) \\
\{ q_\alpha, q_\beta \} &=& \delta_{\alpha \beta} P^+ \\
\eeas
where 
\beas
q &=& {1 \over \sqrt{R}} \tr(\Psi)\\
Q &=& \sqrt{R} \tr(\Pi^a \gamma^a \Psi + \Pi^i \gamma^i \Psi - {i \over 2} [X^i, 
X^j] 
\gamma^{ij} \Psi - {i \over 2} [X^a, X^b] \gamma^{ab} \Psi -i [X^i, X^a] 
\gamma^{ia} \Psi\\&& -{\mu \over 3R} X^i \gamma^{123} \gamma^i \Psi + {\mu 
\over 6R} X^a \gamma^{123} \gamma^a \Psi )\\
\eeas

\section{Perturbative calculations}

In this appendix, we perform explicitly the calculation of the energy shift for 
the lowest non-trivial excited states above the $X=0$ vacuum and the irreducible 
vacuum.

\subsection{$X=0$ vacuum}

Writing out the cubic interaction in terms of the matrix creation and 
annihilation operators, we find 
\beas
H_3 &=& \left({R \over \mu} \right)^{3 \over 2} \tr \left( \sqrt{3 \over 8} i 
\epsilon^{ijk}(A_i A_j A_k + 3 A_i^\dagger A_j A_k + 3 A_i^\dagger A_j^\dagger 
A_k + A_i^\dagger A_j^\dagger A_k^\dagger) \right.\\
&& \qquad - \sqrt{3 \over 2} \psi^{\dagger \alpha} \sigma^i_\alpha {}^\beta [A_i 
+ A_i^\dagger, \psi_\beta] \\
&& \qquad \left. - {\sqrt{3} \over 2} \epsilon_{\alpha \beta} \psi^{\dagger 
\alpha} {\sf g}^a [A_a + A_a^\dagger, \psi^{\dagger \beta}] + {\sqrt{3} \over 
2} \epsilon^{\alpha \beta} \psi_\alpha {\sf g}^{\dagger a} [A_a + A_a^\dagger, 
\psi_\beta] \right)
\eeas
For the quartic term, it will be more convenient to keep the coordinate notation
\beas
H_4 =&& -\left({ R \over \mu}\right)^3 \tr \left( {1 \over 4} [X^i, X^j]^2 + {1 
\over 2} [X^i, X^a]^2 + {1 \over 4} [X^a, X^b]^2  \right)
\eeas
It is easy to see that the energy shift at first order in perturbation theory
\[
\langle \psi | H_3 | \psi \rangle
\]
will be zero for all states, since all terms in $H_3$ have non-zero $H_2$ 
eigenvalue. Thus $H_3$ acting on any state gives a combination of basis state 
all of whose energies are different than that of the original state.

At second order in perturbation theory, the energy shift is given by the usual 
formula,
\bea
\Delta &=& \Delta_4 + \Delta_3 \nonumber\\
&=& \langle \psi | H_4 | \psi \rangle + \sum_n {1 \over E_\psi - E_n} |\langle n 
| H_3 | \psi \rangle |^2 
\label{2pert0}
\eea

We start by verifying that the vacuum state stays at zero energy. The 
contribution to the vacuum shift from $H_4$ is given by
\beas
\Delta_4 &=& \langle 0 | H_4 | 0 \rangle \\
&=& {27 \over 4} (N^3 - N) + 135 (N^3 - N) + 81(N^3 - N)
\eeas
where the three terms come from the $ij$, $ab$, and $ai$ terms in the quartic 
potential respectively.  

To compute $\Delta_3$, note that
\be
\label{three}
H_3 |0 \rangle = \sqrt{3 \over 8} i \sqrt{18(N^3-N)} |A \rangle + \sqrt{3} 
\sqrt{48(N^3 - N)} |B \rangle
\ee
where we have defined orthonormal states
\beas
|A \rangle &=& {1 \over 18(N^3 - N)}\epsilon^{ijk} \tr(A_i^\dagger A_j^\dagger 
A_k^\dagger) |0 \rangle \\
|B \rangle &=& {1 \over 48(N^3 - N)} \epsilon_{\alpha \beta} {\sf 
g}^a_{IJ} 
\tr(\psi^{\dagger I \alpha} \psi^{\dagger J \beta} A_a^\dagger )|0 \rangle 
\eeas
Thus, the contribution to the vacuum shift from $H_3$ is
\beas
\Delta_3 &=& -{1 \over E_A} |\langle A |H_3 | 0 \rangle |^2
           -{1 \over E_B} |\langle B |H_3 | 0 \rangle |^2\\
&=&-{27 \over 4}(N^3 - N) - 216(N^3-N)\\
&=& -\Delta_4
\eeas

Thus, we have shown that the vacuum state remains at zero energy to second order 
in 
perturbation theory. Using (\ref{three}), we may also compute the correction to 
the vacuum wavefunction to first order in perturbation theory, and we find the 
result
\be
\label{vacshift}
|vac \rangle = {1 \over \sqrt{1 + {891 \over 4} \lambda^2 (N^3 - N)}} (|0> -  
\sqrt{3 \over 8} i \lambda \sqrt{18(N^3-N)} |A \rangle -18 \lambda 
\sqrt{(N^3 - N)} |B \rangle )
\ee

We now compute the vacuum shift for the lowest energy non-trivial state (i.e. 
involving more than $U(1)$ oscillators),
\[
|\psi \rangle = {1 \over \sqrt{12(N^2-1)}} (\tr(A_a^\dagger A_a^\dagger)- {1 
\over N} \tr(A_a^\dagger) \tr(A_a^\dagger) ) |0 \rangle 
\]
This normalized state has been chosen to be orthogonal to the state
\[
\tr(A_a^\dagger) \tr(A_a^\dagger) ) |0 \rangle
\]
which cannot receive any energy correction.

To compute $\Delta_3$, we note that $H_3$ may be written as a sum of terms 
$H_3^i$ with specific $H_2$ eigenvalues, 
\[
H_3 = \sum H^i_3 
\] 
such that
\[
[H_2, H^i_3] = E_i H^i_3
\]
where $H_2$ is the quadratic Hamiltonian and all the $E_i$'s are distinct.  

Using this decomposition, we find
\beas
\Delta_3 &=& \sum_n {1 \over E_\psi - E_n} |\langle n | H_3 | \psi \rangle|^2\\
&=& \sum_{n,i,j} -{1 \over E_i} \langle \psi |(H^j_3)^\dagger |n \rangle 
\langle n | H_3^i | \psi \rangle\\
&=& \sum_{n,i} -{1 \over E_i} \langle \psi |(H^i_3)^\dagger |n \rangle 
\langle n | H_3^i | \psi \rangle\\
&=& \sum_i -{1 \over E_i} \langle  \psi |  (H^i_3)^\dagger H^i_3 | \psi \rangle
\eeas
Thus, we may write $\Delta$ as
\be
\label{2pertN}
\Delta = \langle  \psi |  H_4 | \psi \rangle - \sum_i {1 \over E_i} \langle  
\psi |  (H^i_3)^\dagger H^i_3 | \psi \rangle 
\ee

Since the vacuum energy shift was zero, we can ignore all ``disconnected'' 
contributions in which the creation operators from the initial state contract 
only with annihilation operators from the final state since these terms will be 
the same as for the vacuum shift. A further simplification arises from the fact 
that the interaction vertices are written in terms of commutators of $X$'s with 
the result that $\tr(A^\dagger)$ or $\tr(A)$ from the initial or final state 
contracted with any of the interaction terms will vanish. Thus, we may ignore 
the 
\[
- {1 \over N} \tr(A_a^\dagger) \tr(A_a^\dagger)
\]
piece since it contributes only to the disconnected contributions.

The connected contributions to $\Delta_3$ come from the terms
\beas
H^1_3 &=&  \sqrt{3} \epsilon_{\alpha \beta} {\sf g}^a_{IJ} \tr( \psi^{\dagger 
\alpha I} \psi^{\dagger \beta J} A_a)\\
H^2_3 &=& \sqrt{3} \epsilon_{\alpha \beta} {\sf g}^a_{IJ} \tr( \psi^{\dagger 
\alpha I} \psi^{\dagger \beta J} A_a^\dagger)
\eeas
with $H_2$ eigenvalues $E_1 = {1 \over 3}$, and $E_2 = {2 \over 3}$. The two 
contributions to $\Delta_3$ are then
\beas
\Delta_3^1 &=& - 3 \langle  \psi | (H^1_3)^\dagger H^1_3 | \psi \rangle \\
&=& -3 \cdot {1 \over 12(N^2-1)}  \cdot 3 |2 \epsilon_{\alpha \beta} {\sf 
g}^a_{IJ} \tr( \psi^{\dagger \alpha I} \psi^{\dagger \beta J} A_a)|0 \rangle |^2 
\\
&=& -144 N
\eeas
and
\beas 
\Delta_3^2 &=& - {3 \over 2} \langle  \psi | (H^2_3)^\dagger H^2_3 | \psi 
\rangle\\
&=& - {3 \over 2} \cdot {1 \over 12(N^2-1)} \cdot 3 | \epsilon_{\alpha \beta} 
{\sf g}^a_{IJ} \tr( \psi^{\dagger \alpha I} \psi^{\dagger \beta J} A_a^\dagger) 
\tr(A_b^\dagger A_b^\dagger)|^2\\
&=& -72 N
\eeas

The connected part of $\Delta_4$ also receives two contributions, coming from 
\[
H_4^1 = - {1 \over 2} \tr(X^a X^b X^a X^b - X^a X^a X^b X^b)
\]
and
\[
H_4^2 = - \tr(X^a X^i X^a X^i - X^a X^a X^i X^i) 
\]
The computation may be carried out efficiently using diagrams with double-line 
notation, making use of the propagators
\beas
\langle X^i_{kl} X^j_{mn} \rangle &=& {3 \over 2} \delta^{ij} \delta_{kn} 
\delta_{lm} \\
\langle X^a_{kl} X^b_{mn} \rangle &=& 3 \delta^{ab} \delta_{kn} \delta_{lm}\\
\langle X^i_{kl} A^j_{mn} \rangle &=& \langle (A^\dagger)^i_{kl} X^j_{mn} 
\rangle = \sqrt{3 \over 2} \delta^{ij} \delta_{kn} \delta_{lm} \\
\langle X^a_{kl} A^b_{mn} \rangle &=& \langle (A^\dagger)^a_{kl} X^b_{mn} 
\rangle = \sqrt{3} \delta^{ab} \delta_{kn} \delta_{lm} \\
\eeas
Evaluating the various figure eight diagrams we find
\beas
\Delta_4^1 &=& -{1 \over 12(N^2-1)} \cdot  {1 \over 2} \langle \tr(A_c A_c)   
\tr(X^a X^b X^a X^b - X^a X^a X^b X^b) \tr(A_d^\dagger A_d^\dagger) \rangle\\
&=& 270N
\eeas
and 
\beas
\Delta_4^2 &=& -{1 \over 12(N^2-1)} \langle \tr(A_c A_c)   
\tr(X^a X^i X^a X^i - X^a X^a X^i X^i) \tr(A_d^\dagger A_d^\dagger) \rangle\\
&=& 54N
\eeas
Thus, we find that the total energy shift for the lowest nontrivial excited 
state $|\psi \rangle$ is given by
\beas
\Delta &=& \Delta_3^1 + \Delta_3^2 + \Delta_4^1 + \Delta_4^2\\
&=& 108 N \mu \left( {R \over \mu} \right)^3  
\eeas
Thus, the effective coupling controlling the perturbation expansion appears to 
be (ignoring numerical factors)
\[
\lambda_{X=0} = N (R/ \mu)^3 = g^2 N
\] 
where $g = (R / \mu)^{3 \over 2}$ was the bare coupling suppressing the 
interaction terms in the action near $X=0$.

\subsection{Irreducible vacuum}

We now compute the energy shift for an arbitrary occupation number of
the breathing mode of the membrane sphere for the irreducible vacuum, 
\[
| \psi \rangle = {1 \over \sqrt{n!}} (a^\dagger_{00})^n |0 \rangle \; .
\] 
We want to calculate the energy shift upon turning on the interaction terms
\bea
H_3 &=& ({R \over \mu})^{3\over 2}  \tr \left( -{1 \over 3} [J^i, X^a][Y^i, 
X^a] - {1 \over 3} [J^i, Y^j][Y^i, Y^j] + { i \over 3} \epsilon^{ijk} Y^i Y^j 
Y^k 
\right. 
\cr
 & &  
\qquad  \left. + \psi^{\dagger} \sigma^i [Y^i, \psi] 
- {1 \over 2} \epsilon \psi^{\dagger} {\sf g}^a 
[X^a, \psi^{\dagger}] + {1 \over 2} \epsilon \psi ({\sf g}^\dagger)^{a} [X^a, 
\psi] \right) \cr   
H_4 &=& ({R \over \mu})^3 \tr \left( -{ 1 \over 4} [Y^i , Y^j]^2 
- {1 \over 4} [X^a, X^b]^2 - { 1 \over 2} [Y^i , X^a]^2 \right)
\eea

As in our calculation for the irreducible vacuum, the leading correction to the 
energy will come at second order in perturbation theory and will be given by 
\be
\label{2pert2}
\Delta = \langle  \psi |  H_4 | \psi \rangle - \sum_i {1 \over E_i} \langle  
\psi |  (H^i_3)^\dagger H^i_3 | \psi \rangle 
\ee
where, as discussed in the previous section, we have expanded 
\[
H_3 = \sum H^i_3 
\] 
such that
\[
[H_2, H^i_3] = E_i H^i_3
\]
where $H_2$ is the quadratic Hamiltonian and all the $E_i$'s are distinct.  

Again, we need only evaluate the connected contributions, since the disconnected 
pieces will be the same as for the vacuum state and therefore cancel. For all 
connected contributions, the $a$s and $a^\dagger$s from the initial
and final state 
must contract with some $Y$s in $H_3$ or $H_4$, and these Wick contractions may 
be 
performed using the relations
\[
\langle a Y^i_{mn} \rangle = \langle Y^i_{mn} a^\dagger \rangle = \sqrt{6 \over 
N(N^2-1)} J^i_{mn}
\]
In other words, Wick contracting a $Y$ with $a$ or $a^\dagger$ replaces it with 
a $J$. 

The connected contributions may be divided as
\[
\Delta = \tilde{\Delta} + \bar{\Delta}
\]
where $\tilde{\Delta}$ denotes the contributions for which all
but one of the $a$s from the final state contract with $a^\dagger$s
from the initial state, while $\bar{\Delta}$ denotes those for which
all but two of the $a$s from the final state contract with
$a^\dagger$s from the initial state. Taking into account the
normalization factor for the state and the combinatorical factors
associated with pairing $a$s from the final state with $a^\dagger$s
from the initial state, we find
\bea
\tilde{\Delta} &=& n  \cdot \tilde{\Delta}\{ n=1 \} \cr
\bar{\Delta} &=& \left( \ba{c} n \\ 2 \ea \right) \bar{\Delta} \{ n=2
\}  \label{genn}
\eea
We begin by calculating $\tilde{\Delta}\{ n=1 \}$.

The $H_4$ term in $\tilde{\Delta}$ will involve terms in the 
quartic action with two $Y$'s replaced by $J$'s, while the $H_3 
H_3$ term in $\tilde{\Delta}$ will involve terms in the cubic action with one 
$Y$ 
replaced by a $J$. The calculation therefore reduces to evaluating correlation 
functions involving the following operators which are quadratic in the fields.
\beas
A &=& \tr(i \epsilon^{ijk} J^i [Y^j, Y^k]) \\
B &=& \tr([J^i, Y^j]^2) \\
C &=& \tr([J^i, X^a]^2) \\
D &=& \tr(\psi^\dagger \sigma^i [J^i, \psi] + \psi \sigma^i [J^i, \psi^\dagger])
\eeas
In terms of the normalized coordinates, these are given by
\beas
A &=& \sum (j-1) \beta^\dagger_{jm} \beta_{jm} - \sum (j+2) \alpha^\dagger_{jm} 
\alpha_{jm} \\
B &=& - \sum  j(j-1) \beta^\dagger_{jm} \beta_{jm} - \sum (j+1)(j+2) 
\alpha^\dagger_{jm} \alpha_{jm} \\
C &=& - \sum j(j+1) (x^a_{jm})^\dagger x^a_{jm}\\
D &=& \sum (j- {1 \over 2})  (\eta^\dagger_{jm} \eta_{jm} - \eta_{jm} 
\eta^\dagger_{jm}) + \sum (j+ {3 \over 2})  (\chi^\dagger_{jm} \chi_{jm} - 
\chi_{jm} \chi^\dagger_{jm})
\eeas

The expectation values of these operators may be calculated to be
\beas
\langle A \rangle &=& {3 (N-1) \over 2 N}\\
\langle B \rangle &=& - {1 \over 2} (N-1) (4 N^2 + 4 N - 3)\\
\langle C \rangle &=& -6N(N^2-1)\\
\langle D \rangle &=& -{16 \over 3} N (N^2 - 1)
\eeas
using the propagators
\beas
\langle (x^a_{jm})^\dagger x^b_{\tilde{j} \tilde{m}}  \rangle &=& 
\delta^{ab} 
\delta_{j,\tilde{j}} \delta_{m,\tilde{m}} {3 \over 2j + 1}\\
\langle (\alpha_{jm})^\dagger \alpha_{\tilde{j} \tilde{m}}  \rangle &=& 
\delta_{j,\tilde{j}} \delta_{m,\tilde{m}} {3 \over 2(j + 1)}\\
\langle (\beta_{jm})^\dagger \beta_{\tilde{j} \tilde{m}}  \rangle &=& 
\delta_{j,\tilde{j}} \delta_{m,\tilde{m}} {3 \over 2j}\\
\eeas 

The $H_4$ contribution may now be quickly evaluated. We find
\beas
\tilde{\Delta}_4^{XYXY} = \langle a (- {1 \over 2} \tr([X^a,Y^i]^2) ) a^\dagger 
\rangle 
&=& -{1 \over 2} \cdot 2 \cdot {6 \over N (N^2-1)} \langle C \rangle\\
&=& 36
\eeas
while
\beas
\tilde{\Delta}_4^{YYYY} = \langle a (- {1 \over 4} \tr([Y^i,Y^j]^2) ) a^\dagger 
\rangle 
&=& -{1 \over 4} \cdot {6 \over N(N^2-1) }( 8 \langle A \rangle + 4 \langle B 
\rangle)\\
&=& 12 - 9 {N+2 \over N^2 (N+1)}
\eeas
Note that in the first step we have used the fact that 
\[
\tr([J^i,Y^j][Y^i,J^j]) = \tr([J^i,Y^i]^2) + \tr([J^i, J^j][Y^i,Y^j])
\]
but the first term on the right involves only the gauge degrees of freedom, 
so we may set it to zero.

We next calculate
\[
\tilde{\Delta}_3 = -\sum_i {1 \over E_i} \langle 0| a (H^i_3)^\dagger H^i_3  
a^\dagger | 
0 \rangle
\]
The connected contributions are given by
\bea
\tilde{\Delta}_3 &=& -\sum_i {1 \over E_i} \langle 0| [[a,
(H^i_3)^\dagger], a^\dagger] \;  H^i_3 | 0 \rangle 
- \sum_i {1 \over E_i} \langle 0| (H^i_3)^\dagger \; [[a, H^i_3
],a^\dagger] | 0 \rangle \cr
&& -\sum_i {1 \over E_i} \langle 0| [a, (H^i_3)^\dagger] \; [ H^i_3  , 
a^\dagger] | 0 \rangle - \sum_i {1 \over E_i} \langle 0| [(H^i_3)^\dagger , 
a^\dagger] \; [a, H^i_3 ]| 0 \rangle \label{d3}
\eea
To calculate the terms on the second line, note that since $H_3$
depends on $a$ and $a^\dagger$ only in the combination $(a + a^\dagger)$, we 
have
\[
[H_3, a^\dagger] = [a, H_3] \equiv \sum {\cal O}_{E_i}
\]
where we have expanded the left hand expressions into a set of terms with 
distinct $H_2$ eigenvalues. Then
\[
[a, H_3^i] = {\cal O}_{E_i - E_a} \qquad [H^i_3, a^\dagger] 
= {\cal O}_{E_i + E_\a}
\]
Using these expressions, we have
\beas
\Delta_3 &=& -\sum_i {1 \over E_i} \langle 0| {\cal O}^\dagger_{E_i + E_\a} 
{\cal O}_{E_i + E_\a}| 0 \rangle - \sum_i {1 \over E_i} \langle 0| {\cal 
O}^\dagger_{E_i - E_\a}{\cal O}_{E_i - E_\a}| 0 \rangle \\
&=& -\sum_i {2 E_i \over E^2_i - E^2_a} \langle 0| {\cal O}^\dagger_{E_i} {\cal 
O}_{E_i}| 0 \rangle 
\eeas
where $E_a = {1 \over 3}$ is the energy of the oscillator under consideration.

To proceed, we note first that all disconnected terms in the expression above
vanish since these will have $E^i = 0$. To calculate the connected terms, we may 
use the explicit expressions for the operators ${\cal O}$ obtained from 
\beas
[H_3, a^\dagger] &=& \sqrt{6 \over N(N^2-1)} (-{1 \over 6} A - {1 \over 3} B - 
{1 \over 3} C + {1 \over 2} D )\\
 &=& \sqrt{6 \over N(N^2-1)}( {1 \over 6} (j-1) (2j-1) \beta_{jm} 
\beta^\dagger_{jm}\\ && \qquad  +  {1 \over 6} (j+2) (2j+3) \alpha_{jm} 
\alpha^\dagger_{jm} +  {1 \over 3} j (j+1)  x^a_{jm} (x^a_{jm})^\dagger\\ 
&& \qquad + {1 \over 2} (j - {1 \over 2} ) (\eta^\dagger_{jm} \eta_{jm} - 
\eta_{jm} \eta^\dagger_{jm})  + {1 \over 2} (j + {3 \over 2} ) 
(\chi^\dagger_{jm} \chi_{jm} - \chi_{jm} \chi^\dagger_{jm} ) )
\eeas 
Note that the fermions terms contain only pieces with $E_0 = 0$ and therefore 
will not contribute. The $\alpha$, $\beta$, and $x$ contributions may be 
calculated separately using the propagators given above. For example, the 
$\beta$ term will be
\beas
&& {6 \over N(N^2-1)}{-2({2j \over 3}) \over ({2j \over 3})^2 - ({1 \over 
3})^2}\langle {1 \over 6} (j-1)(2j-1) \beta_{jm} \beta^\dagger_{jm} \; {1 \over 
6} (j-1) (2j-1) \beta_{jm} \beta^\dagger_{jm} \rangle \\
= &&  {6 \over N(N^2-1)}  \sum_{j=1}^N  {-2({2j \over 3}) \over ({2j \over 3})^2 
- ({1 \over 3})^2} \cdot ({1 \over 6} (j-1) (2j-1))^2 \cdot 2 \cdot ({3 \over 
2j})^2 (2j+1)\\
= && - {9 \over N(N^2-1)} \sum_{j=1}^N {(j-1)^2 (2j-1) \over j}
\eeas
where in the second line the third factor is from the two ways of contracting 
the $\beta$s, the fourth factor is from the propagators, and the final factor is 
from the sum over $m$.

Similarly, the $\alpha$ contribution gives
\[
- {9 \over N(N^2-1)} \sum_{j=0}^{N-2} {(j+2)^2 (2j+3) \over (j+1)} 
\]
Together, these sums may be evaluated to give
\[
\tilde{\Delta}_3^{YYY} = - 12 - 9 {5N + 1 \over N^2 (N+1)}
\]
The contribution from the $x$ terms is 
\[
\tilde{\Delta}_3^{XXY} = - {108 \over N(N^2-1)} \sum_{j=0}^{N-1} j(j+1) = -36
\]
The remaining contribution to $\tilde{\Delta}_3$ comes from the terms
on the first line of equation (\ref{d3}), which we denote by 
$\tilde{\Delta}'_3$. To compute these, we use
\be
[[a,H_3],a^\dagger] = 3 \sqrt{6 \over N (N^2-1)} (a+a^\dagger)
\ee
Then the first line of equation (\ref{d3}) becomes
\beas
\tilde{\Delta}'_3 &=&-3 \cdot 3 \sqrt{6 \over N (N^2-1)} (\langle 0| a
H_3 | 0 \rangle + \langle 0| H^3 a^\dagger | 0 \rangle ) \\
&=& -{108 \over N(N^2-1)} \langle -{1 \over 6} A - {1 \over 3} B - 
{1 \over 3} C + {1 \over 2} D  \rangle \\
&=& 27(2N+1) \over N^2 (N+1)
\eeas
Combining all the terms in $\tilde{\Delta}$, we see that 
$\tilde{\Delta}_3^{XXY}$ cancels $\tilde{\Delta}_4^{XYXY}$, while 
$\tilde{\Delta}_4^{YYYY}$, $\tilde{\Delta}_3^{YYY}$, and
$\tilde{\Delta}'_3$ sum to zero, so that
\[
\tilde{\Delta} = 0
\]

We now proceed to calculate $\bar{\Delta} \{n = 2\}$. In this case,
the $H_4$ contribution involves only the $Y^4$ terms with each $Y$
contracted with and $a$ or $a^\dagger$. Explicitly, we find
\beas
\bar{\Delta}_4 &=& {1\over 2} \langle a a \; \tr(-{1 \over 4}
[Y^i, Y^j][Y^i,Y^j]) \;  a^\dagger a^\dagger |0 \rangle_* \\
&=& -{1 \over 2} {1 \over 4} \left[{6 \over N (N^2-1)}\right]^2\ 24 
\tr([J^i,J^j]^2)
\\
&=& {54 \over N(N^2-1)}
\eeas
where the $\langle\ \cdot\ \rangle_*$ indicates that we are keeping only 
terms 
with no
contractions between initial and final $a$s. To compute the $H_3 H_3$
terms, we note that 
\[
H_3 = {2 \over 3 \sqrt{N(N^2-1)}} \alpha_{00}^3 + \dots
\] 
where the additional terms have no more than one power of $\alpha_{00}$ and
therefore cannot contribute to $\bar{\Delta}$. Then
\beas
\bar{\Delta}_3 &=&  \sum -{1 \over E_i} \langle \psi | (H_3^i)^\dagger
H_3^i | \psi \rangle_* \\ 
&=& - \sum \langle a a \; \alpha_{00}^3 \;  \alpha_{00}^3 \; a^\dagger a^\dagger
\rangle {4 \over 9 N(N^2-1)} {1 \over 2} \\
&=& {1 \over 2} {4 \over 9 N(N^2-1)} \left\{ \left(\sqrt{3 \over 2}\right)^4 {3 
\over 2}
\right\} (3 \cdot [36] -1 \cdot [36] -3 \cdot [144])\\
&=& {-270 \over N(N^2-1)}
\eeas
where in the last line, terms in $\{ \}$ brackets are from
propagators, and the terms in $[]$ brackets are combinatorical
factors for the contributions with $E_i = -1/3$, $E_i = 1/3$ , and
$E_i = 1$ respectively. 

Thus, for $n=1$ the total energy shift is zero and for $n=2$ we find 
\beas
\bar{\Delta} &=& \bar{\Delta}_3 + \bar{\Delta}_4\\
&=&-{216 \over N(N^1-1)} \; .
\eeas
We may now use (\ref{genn}) to write the result for the leading energy shift
of the breathing mode with occupation number $n$, 
\beas
\Delta &=& - 216 \left( \ba{c} n \\ 2 \ea \right) {R^3 \over \mu^3 N(N^2-1)} \mu 
\\
&=& -216  \left( \ba{c} n \\ 2 \ea \right)  g^2 \mu (1 + {\cal O} ({1 \over N}) 
) 
\eeas
Thus, for the low-energy states, the leading perturbative correction
is proportional to the bare coupling $g^2 = \{R / (\mu N) \}^3$, which
remains finite in the large $N$ limit.

\section{Height of the potential}

In this appendix, we will demonstrate explicitly the existence of a 
path between the irreducible vacuum (labeled by $N+1$) and the vacuum labeled 
by $(N,1)$ along which the maximum energy is of order $(\mu N /  R)^3 {1 
\over N} = {1 \over g^2 N}$ rather than $1/g^2$.\footnote{In terms of the 
natural metric $\tr((X^A)^2)$ on the space of matrices, this choice for the 
final state represents the closest vacuum.}

Our choice of path is physically motivated by the following picture. The initial 
vacuum state corresponds to a single membrane sphere of radius ${\mu N \over 
6R}$, while the final state corresponds to a slightly smaller sphere of radius 
${\mu (N-1) \over 6R}$ plus a small sphere of membrane at the origin. If these 
were really continuum membranes, one way to move between these two points 
continuously at a seemingly small cost in energy would be to form a thin tube of 
membrane from the south pole of the original sphere to the origin
where the small second sphere could 
grow from the tip of the tube. In this way, some of the membrane from the outer 
sphere could be transferred to the origin, yielding the desired final 
configuration, as shown in figure 2.

We can translate this continuum picture to a path between the desired initial 
and 
final matrix configurations as follows. First, write down a map from the unit 
sphere (the membrane worldvolume) to the target space which is a function of 
some parameter $\epsilon$ which will be related to the radius of the inner 
sphere. At a given value of $\epsilon$ there will be some line of latitude near 
the south pole of the worldvolume sphere across which there is a singularity 
corresponding to the infinitesimal tube. The area above this line will map to 
the outer sphere, while the (small) area below this line will map to the small 
inner sphere.\footnote{We will not need to explicitly specify the mapping, but a 
natural choice would be to take the sum of the radii of the target space spheres 
to be a constant, and impose the condition that the map preserves ratios of 
areas.}

Once we have defined such a continuum mapping, we may obtain a corresponding 
path in the space of matrices by expanding the continuum map in terms of 
spherical harmonics,
\[
x^i(\theta, \psi, \epsilon) = \sum x^i_{lm}(\epsilon) Y_{lm}(\theta, \phi)
\] 
and then replacing the spherical harmonics with matrix spherical harmonics with 
$l<N$,
\[
X^i(\epsilon) = \sum_{l < N} x^i_{lm} (\epsilon) Y^{N \times N}_{lm} \; .
\]

Without explicitly specifying a map, it is clear that a natural choice is to 
preserve axial symmetry at all points along the path, so that we may choose 
$x^3$ to be a function of $\theta$ only and $x^1 + i x^2 = r(\theta) e^{i 
\phi}$. In this case, the expansion of $x^3(\theta, \phi)$ will involve only 
spherical harmonics with $m=0$, while the expansion of $x^1 + i x^2$ will 
involve only spherical harmonics with $m=1$. As a result, at every point along 
the path, our matrices will take the form
\[
X^3 = {\mu \over 3R} \left( \ba{cccc} a_1 & & & \\ & a_2 & & \\ & & \ddots & \\ 
& & & a_{N+1} \ea \right) \qquad X^+ = {\mu \over 3R} \left( \ba{cccc} 0 & b_1 & 
& \\ & 0 & \ddots & \\ & & \ddots & b_N \\ & & & 0 \ea \right)  
\]
since the matrices $Y_{l0}$ are all diagonal, while the matrices $Y_{l1}$ have 
nonzero entries only in the first row above the diagonal in each column. 

It is convenient to write the potential explicitly in terms of these variables. 
We find
\be
\label{abpotl}
V = {1 \over 2} \mu \left( {\mu \over 3R}\right)^4 
\left( \sum_{l=1}^{N+1} (a_l + {1 \over 
2}(b_{l-1}^2 - b_l^2))^2 + \sum_{l=1}^N b_l^2 ( a_{l+1} - a_l + 1)^2 
\right) \ee
Since this expression is quadratic in the variables $b_l^2$, it is not hard to 
determine the set of $b$'s for which the potential is minimum for a given set of 
$a$'s. The extremum condition for $b_l$ is
\[
2b_l \left( (a_{i+1} - a_i + {3 \over 2} )^2 - {5 \over 4} + b_l^2 - {1 \over 
2}(b_{l+1}^2 + b_{l-1}^2) \right) = 0
\]  
Thus, at the minimum of the potential for a fixed set of $a$'s, we will have for 
each $l$ either $b_l$=0 or
\be
\label{extremum}
b_{l+1}^2 - 2 b_l^2 + b_{l-1}^2 = 2(a_{l+1} - a_l + {3 \over 2})^2 - {5 \over 2}
\ee

Our problem is now reduced to specifying $a_l(\epsilon)$ such that the initial 
configuration is 
\[
a_l(0) = ({N \over 2}, {N \over 2} - 1, \dots , -{N \over 2} + 1, - {N \over 2} 
)  
\]
and the final configuration is
\[
a_l(1) = ({N -1 \over 2},{N - 3 \over 2}, \dots, -{N-1 \over 2}, 0)
\]
In principle, one could let the value of $a_N$ parameterize the path and 
explicitly extremize the potential in terms of the other $a_i$'s to determine 
the lowest energy path satisfying our ansatz of axial symmetry. However, since 
the values of $a_l$ for $l \le N$ change very little between their initial and 
final values it seems likely that this minimum path will have a maximum energy 
that does not differ significantly from that of the linear path between initial 
and final points. In other words, we expect that working to linear order in 
$\epsilon$ will be sufficient to determine the order of magnitude of the barrier 
height.

Thus, we take
\[
a_l = {N \over 2} + 1 - l - {\epsilon \over 2} \qquad a_{N+1} = - {N \over 2} + 
{N \over 2} \epsilon
\] 
At $\epsilon=0$, the values of $b$ correspond to $X^+ = {\mu \over 3R} J^+$,
\[
b_k^2 = k(N+1 - k) 
\]
so near $\epsilon=0$ we will be on the branch with all $b_i$'s nonzero. 
Inserting the values of $a$ into (\ref{extremum}), we find
\[
\left( \ba{cccc} -2 & 1 & & \\1 & -2 &  & \\ & & \ddots & 1 \\ & & 1 & -2 \ea 
\right) \left( \ba{c} b_1^2 \\ \vdots \\ b_{N-1}^2 \\ b_N^2 \ea \right) = \left( 
\ba{c} -2 \\ \vdots \\ -2 \\ -2 + {1 \over 2} (N+1)^2 \epsilon^2 + (N+1)\epsilon
 \ea \right)
\]  
The matrix $M$ on the left may be inverted to give a symmetric matrix with
\[
M^{-1}_{kl} = -k { N+1 -l \over N+1} \qquad k \le l 
\]
so we find that near $\epsilon = 0$ 
\[
b_k^2 = k(N+1-k -\epsilon - {1 \over 2} (N+1) \epsilon^2)
\]
Note that $b_N$ in this formula is negative for $\epsilon > (\sqrt{2N+3} - 
1)/(N+1)$, so for these values of $\epsilon$, we are no longer on the branch 
with all $b$'s non-zero. In fact, at $\epsilon = (\sqrt{2N+3} - 1)/(N+1)$ the 
values of $b$, 
\[
b_k^2 = k(N-k) 
\]
are already equal to their $\epsilon = 1$ values corresponding to $J^+$ in the 
$(N,1)$ representation. It is not hard to check that these values, together 
with 
the values of $a$ defined above, satisfy the equations (\ref{extremum}) 
for all
values of $\epsilon$. Thus, we may conclude that
\[
b_k^2 = \left\{ \ba{ll}  k(N+1-k -\epsilon - {1 \over 2} (N+1) \epsilon^2) & 
\epsilon \le {\sqrt{2N+3} - 1 \over N+1}\\
k(N-k) & \epsilon \ge {\sqrt{2N+3} - 1 \over N+1} \\ 
\ea \right.
\]
Plugging the values for $a$ and $b$ into (\ref{abpotl}) we conclude that the 
energy along our path is given by
\[
V = \left\{ \ba{ll}  \mu \left( {\mu N \over R} \right)^3 {1 \over 648} 
\left({(N+1) \epsilon \over N} \right)^2 (1 - \epsilon - {1 \over 4} (N+1) 
\epsilon^2) & \epsilon \le {\sqrt{2N+3} - 1 \over N+1}\\
\mu \left( {\mu N \over R} \right)^3 {1 \over 648}{(N+1) \over N^2} (1- 
\epsilon)^2 & \epsilon \ge {\sqrt{2N+3} - 1 \over N+1}\\ 
\ea \right.
\]
From this expression, we find that the maximum of the potential along the path 
we have determined is 
\[
V_{max}/\mu = \left( {\mu N \over R}\right)^3 \left\{ {1 \over 648 N} + 
{\cal O} 
\left({1 \over N^{3 \over 2}} \right) \right\}
\]
This represents an upper bound on the classical energy required to move between 
the vacuum labeled by $\{ N+1 \}$ and the vacuum labeled by $\{N, 1 \}$. It is 
clear that the same bound applies for a path between vacua $\{N+1, N_1, \dots , 
N_n \}$ and $\{1, N, N_1, \dots, N_n\}$. Thus, the minimum energy required to 
move from a given vacuum $\{N_1, N_2, \dots, N_n \}$ to another vacuum is not 
more than ${1 \over N_{min}} (\mu N_{min} / R)^3$, where $N_{min}$
corresponds to the smallest sphere in the collection.

Finally, we note that by a series of these steps in which we move from a vacuum 
$\{N_1, N_2, \dots, N_n\}$ to a vacuum $\{1, N_1 - 1, N_2, \dots , N_n \}$, and 
then by the reverse process to the vacuum $\{N_1, N_2 + 1, \dots, N_n \}$, we 
may transfer any number of momentum units between any two membrane spheres and 
in this way move between any two vacua of the theory. The maximum energy 
required for a given initial and final vacuum will be of order
\[
E \sim {1 \over N_{max}} \left( {\mu N_{max} \over R} \right)^3
\]
where $N_{max}$ corresponds to the largest sphere whose momentum is changed in 
the process.

\end{document}